\documentclass[12pt]{article}
\usepackage{amssymb, amsmath, amsfonts}

\newtheorem{lemma}{Lemma}

\newtheorem{cor}{Corollary}
\newtheorem{theorem}{Theorem}

\begin{document}

\begin{center}
{\Large On the Classification of Darboux Integrable Chains}

\vskip 0.2cm

{Ismagil Habibullin}\footnote{e-mail: habibullin\_i@mail.rb.ru,
(On leave from Ufa Institute of Mathematics, Russian Academy of
Science, Chernyshevskii Str. , 112, Ufa, 450077, Russia)}

{Natalya Zheltukhina}

{Asl{\i} Pekcan}

{Department of Mathematics, Faculty of Science,
 \\Bilkent University, 06800, Ankara, Turkey \\}

\end{center}

\begin{abstract}
We study differential-difference equation of the form
$t_{x}(n+1)=f(t(n),t(n+1),t_x(n))$ with unknown $t=t(n,x)$
depending on $x$, $n$. The equation is called Darboux integrable,
if there exist  functions  $F$  (called an $x$-integral) and $I$
(called an $n$-integral), both  of a finite number of variables
$x$, $t(n)$, $t(n\pm 1)$, $t(n\pm 2)$, $\ldots$, $t_x(n)$,
$t_{xx}(n)$, $\ldots$, such that $D_xF=0$ and $DI=I$, where $D_x$
is the operator of total differentiation with respect to $x$, and
$D$ is the shift operator: $Dp(n)=p(n+1)$. The Darboux
integrability property is reformulated in terms of characteristic
Lie algebras that gives an effective tool for classification of
integrable equations. The complete list of equations of the form
above admitting nontrivial $x$-integrals is given in the case when
the function $f$ is of the special form $f(x,y,z)=z+d(x,y)$.
\end{abstract}


\section{Introduction}

In this paper we  study integrable semi-discrete chains of the
following form
\begin{equation}\label{dhyp}
t_{x}(n+1)=f(t(n),t(n+1),t_x(n)),
\end{equation}
where the unknown $t=t(n,x)$ is a function of two independent
variables: discrete $n$ and continuous $x$. Chain (\ref{dhyp}) can
also be interpreted as an infinite system of ordinary differential
equations for the sequence of the variables
$\{t(n)\}_{n=-\infty}^{\infty}$. Here $f=f(t,t_1,t_x)$ is
assumed to be locally analytic function of three variables
satisfying at least locally the condition
\begin{equation}\label{nonzero}
\frac{\partial f}{\partial t_x}\neq 0.
\end{equation}
For the sake of convenience we introduce subindex denoting shifts
$t_k=t(n+k,x)$ (keep $t_0=t$) and derivatives
$t_x=\displaystyle{\frac{\partial}{\partial x}}t(n,x),$
$t_{xx}=\displaystyle{\frac{\partial^2}{\partial x^2}}t(n,x)$, and
so on. We denote through $D$ and $D_x$ the shift operator and,
correspondingly, the operator of total derivative with respect to
$x$. For instance, $Dh(n,x)=h(n+1,x)$ and
$D_xh(n,x)=\frac{\partial}{\partial x}h(n,x)$. Set of all the
variables $\{t_k\}_{k=-\infty}^{\infty};$ $\{D_x^m
t\}_{m=1}^{\infty}$ constitutes the set of dynamical variables.
Below we consider the dynamical variables as independent ones.
Since in the literature the term "integrable" has various meanings
let us specify the meaning used in the article. Introduce first
notions of $n$- and $x$-integrals \cite{AdlerStartsev}.

Functions $I$ and $F$, both depending on $x$ and a finite number of
dynamical variables, are called  respectively $n$- and $x$-integrals
of (\ref{dhyp}), if  $DI=I$ and $D_xF=0$.

{\bf Definition}. Chain (\ref{dhyp}) is called integrable (Darboux
integrable) if it admits  a nontrivial $n$-integral  and a
nontrivial $x$-integral.

Darboux integrability implies the so-called C-integrability. Knowing
both integrals $F$ and $I$  a Cole-Hopf type differential
substitution $w=F+I$  reduces the equation (\ref{dhyp}) to the
discrete version of D'Alembert wave equation $w_{1x}-w_x=0$. Indeed,
$(D-1)D_x(w)=(D-1)D_xF+D_x(D-1)I=0$.

It is remarkable that an integrable chain is reduced to a pair
consisting of an ordinary differential equation and an ordinary
difference equation. To illustrate it note first that any
$n$-integral might depend only on $x$ and $x$-derivatives of the
variable $t$: $I=I(x,t,t_{x},t_{xx},...)$ and similarly any
$x$-integral depends only on $x$ and the shifts: $F=F(x,t,t_{\pm
1},t_{\pm 2},...)$. Therefore each solution of the integrable chain
(\ref{dhyp}) satisfies two equations:
$$I(x,t,t_{x},t_{xx},...)=p(x), \quad F(x,t,t_{\pm 1},t_{\pm 2},...)=q(n)$$
with properly chosen functions $p(x)$ and $q(n)$.

Nowadays the discrete phenomena are studied intensively due to their
various applications in physics. For the discussions and references
we refer to the articles \cite{AdlerStartsev}, \cite{Zabrodin},
\cite{Yamilov}, \cite{NijhoffCapel}, \cite{GKP}.

Chain (\ref{dhyp}) is very close to a well studied object -- the
partial differential equation of the hyperbolic type
\begin{equation}\label{hyp}
u_{xy}=f(x,y,u,u_x,u_y).
\end{equation}
The definition of integrability  for equation (\ref{hyp}) was
introduced by G. Darboux. The famous  Liouville equation
$u_{xy}=\mathrm{e}^u$ provides an illustrative example of the
Darboux integrable equation. An effective criterion of integrability
of (\ref{hyp}) was discovered by Darboux himself: equation
(\ref{hyp}) is integrable if and only if the Laplace sequence of the
linearized equation terminates at both ends (see \cite{Darboux},
\cite{AndersonKamran}, \cite{Zhiber}). This criterion of
integrability was used in \cite{Zhiber}, where the complete list of
all Darboux integrable  equations of form (\ref{hyp}) is given.

An alternative approach to the classification problem based on the
notion of the characteristic Lie algebra of hyperbolic type systems
was introduced years ago in \cite{ShabatYamilov},
\cite{LeznovSmirnovShabat}. In these articles an algebraic criterion
of Darboux integrability property has been formulated. An important
classification result was obtained in \cite{ShabatYamilov} for the
exponential system
\begin{equation}\label{shabat}
u^i_{xy}=\exp{(a_{i1}u^1+a_{i2}u^2+ ...+a_{in}u^n)}, \quad
i=1,2,...,n.
\end{equation}
It was proved that system (\ref{shabat}) is Darboux integrable if
and only if the matrix $A=(a_{ij})$ is the Cartan matrix of a
semi-simple Lie algebra. Properties of the characteristic Lie
algebras of the hyperbolic systems
\begin{equation}\label{bormisov}
u^i_{xy}=c^i_{jk}u^ju^k, \quad i,j,k=1,2,...,n
\end{equation}
have been studied in \cite{ZhiberMukminov}, \cite{BormisovMukminov}.
Hyperbolic systems of general form admitting integrals are studied
in \cite{SokolovStartsev}. A promising idea of adopting the
characteristic Lie algebras to the problem of classification of the
hyperbolic systems which are integrated by means of the inverse
scattering transforms method is discussed in \cite{zhiberMurtazina}.

The method of characteristic Lie algebras is closely connected with
the symmetry approach \cite{IbragimovShabat} which is proved to be a
very effective tool to classify integrable nonlinear equations of
evolutionary type \cite{MSY}, \cite{LeviYamilov},  \cite{Gurses3},
\cite{Gurses4}, \cite{Svinolupov} (see also the survey
\cite{Yamilov} and references therein). However, the symmetry approach
meets very serious difficulties when applied to hyperbolic type
models. After the papers \cite{ZhiberShabat79} and
\cite{ZhiberShabat84} it became clear that this case needs
alternative methods.

 In this article an algorithm of classification of integrable
 discrete chains of the form (\ref{dhyp}) is suggested based on the notion of
 the characteristic Lie algebra (see also \cite{Habibullin},
\cite{HabibullinPekcan}, \cite{TJM}). Introduce necessary
definitions.

 Define vector fields
\begin{equation}\label{definitionYj}
Y_j=D^{-j}\frac{\partial}{\partial t_1}D^j , \qquad j\geq 1,
\end{equation}
and
\begin{equation}\label{definitionXj}
X_j=\frac{\partial}{\partial_{t_{-j}}}, \qquad  j\geq
1.\end{equation} The following theorem  (see
\cite{HabibullinPekcan})  defines the characteristic Lie algebra
$L_n$ of (\ref{dhyp}).
\begin{theorem}\label{thm1}
Equation (\ref{dhyp}) admits a nontrivial $n$-integral if and only
if the following two conditions hold:\\
1)  Linear space spanned by  the operators $\{Y_j \}_1^{\infty}$ is
of finite dimension, denote this dimension by $N$;\\
2)  Lie algebra $L_n$ generated by the operators
${Y_1,Y_2,...,Y_N,X_1,X_2,...,X_N}$ is of finite dimension. We call
$L_n$ the characteristic Lie algebra of (\ref{dhyp}) in the
direction of $n$.
\end{theorem}

To introduce the characteristic Lie algebra $L_x$ of (\ref{dhyp}) in
the direction of $x$, consider vector fields
\begin{equation}\label{gc1} K_0= \frac{\partial }{\partial x}+t_x\frac{\partial
}{\partial t} +f\frac{\partial }{\partial t_1 }+g\frac{\partial
}{\partial t_{-1}} +f_1\frac{\partial }{\partial
t_2}+g_{-1}\frac{\partial }{\partial t_{-2} }+\ldots\,
\end{equation}
and
\begin{equation}\label{gc1'}
X=\frac{\partial }{\partial t_x}.
\end{equation}
Note that an $x$-integral $F$ solves the equation $K_0F=0$. One
can get this equation by applying the chain rule to the equation
$D_xF=0$, here the function $g$ is defined by the equation
(\ref{dhyp}) rewritten due to (\ref{nonzero}) as
$t_{x}(n-1)=g(t(n),t(n-1),t_x(n))$. Since $F$ does not depend on
the variable $t_x$ one gets $XF=0$. Therefore, any vector field
from the Lie algebra generated by $K_0$ and $X$ annulates $F$.
This algebra is called the characteristic Lie algebra $L_x$ of the
chain (\ref{dhyp}) in the $x$-direction.

The following result is essential,  its proof is a simple
consequence of the famous Jacobi theorem (Jacobi theorem is
discussed, for instance,  in \cite{LeznovSmirnovShabat}).
\begin{theorem}\label{thm1x} Equation
  (\ref{dhyp}) admits a nontrivial $x$-integral if and only if its Lie
algebra $L_x$ is of finite dimension.
\end{theorem}

In the present paper we restrict ourselves to consideration of
existence of $x$-integrals for a particular kind of chain
(\ref{dhyp}), namely, we study chains of the form
\begin{equation}\label{main}
t_{1x}=t_x+d(t,t_1)
\end{equation}
admitting nontrivial $x$-integrals. The main result of the paper,
Theorem \ref{Theoremmain} below, is the complete list of chains
(\ref{main}) admitting nontrivial $x$-integrals.

\begin{theorem}\label{Theoremmain}
Chain (\ref{main}) admits a nontrivial $x$-integral if and only if
$d(t,t_1)$ is one of the kind:
\begin{enumerate}
\item[(1)] $d(t,t_1)=A(t-t_1)$,\\
\item[(2)] $d(t,t_1)=c_0(t-t_1)t+c_2(t-t_1)^2+c_3(t-t_1)$,\\
\item[(3)] $d(t,t_1)=A(t-t_1)\mathrm{e}^{\alpha t}$,\\
\item[(4)] $d(t,t_1)=c_4(\mathrm{e}^{\alpha t_1}-\mathrm{e}^{\alpha t})+
c_5(\mathrm{e}^{-\alpha t_1}-\mathrm{e}^{-\alpha t})$,\\
\end{enumerate}
\noindent where $A=A(t-t_1)$ is a function of $\tau=t-t_1$ and
$c_0, c_2, c_3, c_4, c_5$ are some constants with $c_0\neq 0$,
$c_4\neq 0$, $c_5\neq 0$, and $\alpha$ is a nonzero constant.
Moreover, $x$-integrals in each of the cases are
\begin{enumerate}
\item[i)] $F=x+\int^{\tau}\frac{du}{A(u)}$,\quad if \quad $A(u)\neq0,$\\
$F=t_1-t$,\quad if\quad $A(u)\equiv0$,\\
 \item[ii)]
$F=\frac{1}{(-c_2-c_0)}\ln{|(-c_2-c_0)\frac{\tau_1}{\tau_2}+c_2|}+
\frac{1}{c_2}\ln{|c_2\frac{\tau_1}{\tau}-c_2-c_0|}$\, for $\,c_2(c_2+c_0)\neq0,$\\

$F=\ln{\tau_1}-\ln{\tau_2}+\frac{\tau_1}{\tau}$ for $\,c_2=0,$\\

$F=\frac{\tau_1}{\tau_2}-\ln{\tau}+\ln{\tau_1}$ for $\,c_2=-c_0,$ \\
\item[iii)] $F=\int^{\tau}\frac{\mathrm{e}^{-\alpha u}du}{A(u)}-\int^{\tau_1}\frac{du}{A(u)}$,\\
\item[iv)] $F=\frac{(\mathrm{e}^{\alpha t}-\mathrm{e}^{\alpha
t_2})(\mathrm{e}^{\alpha t_1}-\mathrm{e}^{\alpha
t_3})}{(\mathrm{e}^{\alpha t}-\mathrm{e}^{\alpha
t_3})(\mathrm{e}^{\alpha t_1}-\mathrm{e}^{\alpha t_2})}$.
\end{enumerate}
\end{theorem}

\noindent The $n$-integrals of chain (\ref{main}) can be studied
in a similar way by using Theorem \ref{thm1}, but this problem is
out of the frame of the present article.

The article is organized as follows. In Section  2, by using the
properly chosen sequence of multiple commutators, a very rough
classification result is obtained: function $d(t,t_1)$ for chain
(\ref{main}) admitting $x$-integrals is a quasi-polynomial on $t$
with coefficients depending of $\tau=t-t_1$. Then it is observed
that the exponents $\alpha_0=0$, $\alpha_1$, ..., $\alpha_s$ in
the expansion (\ref{formalpha1}) cannot be arbitrary. For example,
if the coefficient before $\mathrm{e}^{\alpha_0t}=1$ is not
identically zero then the quasi-polynomial $d(t,t_1)$ is really a
polynomial on $t$ with coefficients depending on $\tau$. In
Section 3 we prove that the degree of this polynomial is at most
one. If $d$ contains a term of the form
$\mu(\tau)t^j\mathrm{e}^{\alpha_kt}$ with $\alpha_k\neq0$ then
$j=0$ (Section 4). In Section 5 it is proved that if $d$ contains
terms with $\mathrm{e}^{\alpha_kt}$ and $\mathrm{e}^{\alpha_jt}$
having nonzero exponents then $\alpha_k=-\alpha_j$. This last case
contains chains having infinite dimensional characteristic Lie
algebras for which the sequence of multiple commutators grows very
slowly. They are studied in Sections 6-7. One can find the well
known semi-discrete version of the sine-Gordon (SG) model among
them. It is worth mentioning that in Section 7 the characteristic
Lie algebra $L_x$ for semi-discrete SG is completely described.
The last Section 8 contains the proof of the main Theorem 3 and
here the method of constructing of $x$-integrals is also briefly
discussed.

\section{The first integrability condition}

Define a class \textbf{F} of locally analytic functions each of
which depends only on a finite number of dynamical variables. In
particular we assume that $f(t,t_1,t_x)\in \textbf{F}$. We will
consider vector fields given as infinite formal series of the form
\begin{equation}\label{formal}
Y=\sum_{-\infty}^\infty y_k\frac{\partial}{\partial t_k}
\end{equation}
with coefficients $y_k\in \textbf{F}$. Introduce notions of
linearly dependent and independent sets of the vector fields
(\ref{formal}). Denote through $P_N$ the projection operator
acting according to the rule
\begin{equation}\label{projection}
P_N(Y)=\sum_{k=-N}^{N} y_k\frac{\partial}{\partial t_k}.
\end{equation}
First we consider finite vector fields as
\begin{equation}\label{finitefield}
Z=\sum_{k=-N}^{N}z_k\frac{\partial}{\partial t_k}.
\end{equation}
We say that a set of finite vector fields $Z_1$, $Z_2$, ..., $Z_m$
is linearly dependent in some open region \textbf{U}, if there is
a set of functions $\lambda_1,\,\lambda_2,\, ...,\lambda_m$ defined
on \textbf{U} such that the function
$|\lambda_1|^2+|\lambda_2|^2+...+|\lambda_m|^2$ does not vanish
identically and the condition
\begin{equation}\label{linear combination}
\lambda_1Z_1+\lambda_2Z_2+ ...+\lambda_mZ_m=0
\end{equation}
holds for each point of region \textbf{U}.

We call a set of the  vector fields $Y_1$, $Y_2$, ..., $Y_m$ of the
form (\ref{formal}) linearly dependent in the region \textbf{U} if
for each natural $N$ the following set of finite vector fields
$P_N(Y_1)$, $P_N(Y_2)$, ..., $P_N(Y_m)$ is linearly dependent in this
region. Otherwise we call the set $Y_1$, $Y_2$, ..., $Y_m$ linearly
independent in \textbf{U}.

The following proposition is very useful, its proof is almost
evident.

{\bf Proposition}. { \it If a vector field $Y$ is expressed as a
linear combination
\begin{equation}\label{lincombination}
Y=\lambda_1Y_1+\lambda_2Y_2+ ...+\lambda_mY_m,
\end{equation}
where the set of vector fields $Y_1$, $Y_2$, ..., $Y_m$ is
linearly independent in \textbf{U} and the coefficients of all the
vector fields $Y$, $Y_1$, $Y_2$, ..,. $Y_m$ belonging to
\textbf{F} are defined in \textbf{U} then the coefficients
$\lambda_1,\,\lambda_2,\, ...,\lambda_m$ are in \textbf{F}.}

Below we concentrate on the class of chains of the form
(\ref{main}). For this case the Lie algebra $L_x$ splits down into
a direct sum of two subalgebras. Indeed, since $f=t_x+d$ and
$g=t_x-d_{-1}$ one gets $f_k=t_x+d+\sum_{j=1}^k d_j$ and
$g_{-k}=t_x-\sum_{j=1}^{k+1}d_{-k},$ for $k\geq 1$, where $d=d(t,
t_1)$ and $d_j=d(t_j,t_{j+1})$. Due to this observation the vector
field $K_0$ can be rewritten as $ K_0=t_x\tilde{X}+Y\, , $ with
\begin{equation}\label{tildeX}
\tilde{X}= \frac{\partial }{\partial t}+ \frac{\partial }{\partial
t_{1}}+ \frac{\partial  }{\partial t_{-1} }+ \frac{\partial
}{\partial t_{2} }+  \frac{\partial }{\partial t_{-2} }+\ldots\,
\end{equation}
and
$$
{Y}= \frac{\partial  }{\partial x } +d  \frac{\partial }{\partial
t_{1}}-d_{-1} \frac{\partial  }{\partial t_{-1} }+ (d+d_1)
\frac{\partial }{\partial t_{2} }- (d_{-1}+d_{-2}) \frac{\partial
}{\partial t_{-2} }+\ldots\, .
$$
Due to the relations $[X, \tilde{X}]=0$ and $[X, {Y}]=0$ we have
$\tilde{X}=[X,K_0]\in L_x$, hence ${Y}\in L_x.$ Therefore
$L_x=\{X\}\bigoplus L_{x1},$ where $L_{x1}$ is the Lie algebra
generated by the operators $\tilde{X}$ and ${Y}$.

\begin{lemma} If equation (\ref{main}) admits a nontrivial
$x$-integral then it admits a nontrivial $x$-integral $F$ such that
$\displaystyle{\frac{\partial F}{\partial x}}=0$.
\end{lemma}

\noindent{\bf{Proof}}.  Assume that a nontrivial $x$-integral of
(\ref{main}) exists. Then the Lie algebra $L_{x1}$ is of finite
dimension. One can choose a basis of $L_{x1}$ in the form
$$
\begin{array}{ll}
T_1=\displaystyle{\frac{\partial}{\partial
x}+\sum_{k=-\infty}^\infty
a_{1,k}\frac{\partial }{\partial t_k}}\, ,\\
\\
T_{j}=\displaystyle{\sum_{k=-\infty}^\infty a_{j,k}\frac{\partial
}{\partial t_k}}\, , & 2\leq j\leq N.
\end{array}
$$
Thus, there exists an $x$-integral $F$ depending on $x$, $t$, $t_1$,
$\ldots$, $t_{N-1}$ satisfying the system of equations
$$
\begin{array}{ll}
\displaystyle{\frac{\partial F}{\partial x}+\sum_{k=0}^{N-1}
a_{1,k}\frac{\partial F}{\partial t_k}}=0\, ,\\
\\
\displaystyle{\sum_{k=0}^{N-1} a_{j,k}\frac{\partial  F}{\partial
t_k}}=0\, , & 2\leq j\leq N.
\end{array}
$$
Due to the famous Jacobi Theorem \cite{LeznovSmirnovShabat} there is
a change of variables $\theta_j=\theta_j(t, t_1, \ldots, t_{N-1})$
that reduces the system to the form
$$
\begin{array}{ll}
\displaystyle{\frac{\partial F}{\partial x}+\sum_{k=0}^{N-1}
\tilde{a}_{1,k}\frac{\partial F}{\partial \theta_k}}=0\, ,\\
\\
\displaystyle{\frac{\partial  F}{\partial \theta_k}}=0\, , & 2\leq
j\leq N-2
\end{array}
$$
that is equivalent to
$$
\displaystyle{\frac{\partial F}{\partial x}+
\tilde{a}_{1,N-1}\frac{\partial F}{\partial \theta_{N-1}}}=0
$$
for $F=F(x, \theta_{N-1})$. \\
There are two possibilities: 1) $\tilde{a}_{1, N-1}=0$ and 2)
$\tilde{a}_{1, N-1}\ne 0$. In case 1), we at once have
$\displaystyle\frac{\partial F}{\partial x}=0$. In case 2),
$F=x+H(\theta_{N-1})=x+H(t, t_1,\ldots,  t_{N-1})$ for some
function $H$. Evidently, $F_1=DF=x+H(t_1, t_2, \ldots, t_{N})$ is
also an $x$-integral, and $F_1-F$ is a nontrivial $x$-integral not
depending on $x$. $\Box$

Below we look for $x$-integrals $F$ depending on dynamical variables
$t$, $t_{\pm 1}$, $t_{\pm 2}$, $\ldots$ only (not depending on $x$).
In other words, we study Lie algebra generated by vector fields
$\tilde{X}$ and $\tilde{Y}$, where
\begin{equation}\label{tildeY}
\tilde{Y}= d  \frac{\partial }{\partial t_{1}}-d_{-1} \frac{\partial
}{\partial t_{-1} }+ (d+d_1) \frac{\partial }{\partial t_{2} }-
(d_{-1}+d_{-2}) \frac{\partial }{\partial t_{-2} }+\ldots\, .
\end{equation}
One can prove that the linear operator $Z\rightarrow DZD^{-1}$
defines an automorphism of the characteristic Lie algebra $L_x$.
This automorphism plays the crucial role in all of our further
considerations. Further we refer to it as the shift automorphism.
For instance, direct calculations show that
\begin{equation}
\label{DtildeXD}
 D\tilde{X}D^{-1}=\tilde{X},
\qquad D\tilde{Y}D^{-1}=-d\tilde{X}+\tilde{Y} \, .\end{equation}

\begin{lemma}\label{Lemma0}
Suppose that a vector field of the form $Z=\sum a(j)\frac{\partial
}{\partial t_{j} }$ with the coefficients $a(j)=a(j,t,t_{\pm
1},t_{\pm 2},...)$ depending on a finite number of the dynamical
variables solves an equation of the form $DZD^{-1}=\lambda Z.$ If
for some $j=j_0$ we have $a(j_0)\equiv0$ then $Z=0$.
\end{lemma}

\noindent {\bf Proof}. By applying the shift automorphism to the
vector field $Z$ one gets $DZD^{-1}=\sum D(a(j))\frac{\partial
}{\partial t_{j+1} }$. Now, to complete the proof, we compare the
coefficients of $\frac{\partial }{\partial t_{j} }$ in the
equation $\sum D(a(j))\frac{\partial }{\partial t_{j+1} }=\lambda
\sum a(j)\frac{\partial }{\partial t_{j} }$. $\Box$

Construct an infinite sequence of multiple commutators of the
vector fields $\tilde{X}$ and $\tilde{Y}$
\begin{equation}
\label{tildeY_j} \tilde{Y}_1=[\tilde{X}, \tilde{Y}], \qquad
\tilde{Y}_k=[\tilde{X}, \tilde{Y}_{k-1}] \quad\mbox{for}\quad
k\geq 2\, .
\end{equation}
\begin{lemma}\label{Lemma1}
We have,
\begin{equation}\label{DtildeYkD}
D\tilde{Y}_kD^{-1}=-\tilde{X}^k(d)\tilde{X}+\tilde{Y}_k, \quad
k\geq 1.
\end{equation}
\end{lemma}

\noindent {\bf Proof}. We prove the statement by induction on $k$.
Base of induction holds. Indeed, by (\ref{DtildeXD}) and
(\ref{tildeY_j}), we have
$$
D\tilde{Y}_1D^{-1}=D[\tilde{X},
\tilde{Y}]D^{-1}=[D\tilde{X}D^{-1}, D\tilde{Y}D^{-1}]=[\tilde{X},
-d\tilde{X}+\tilde{Y}]=-\tilde{X}(d)\tilde{X}+\tilde{Y}_1.
$$
Assuming the equation (\ref{DtildeYkD}) holds for $k=n-1$, we have
$$
D\tilde{Y}_nD^{-1}=[D\tilde{X}D^{-1},
D\tilde{Y}_{n-1}D^{-1}]=[\tilde{X},
-\tilde{X}^{n-1}(d)\tilde{X}+\tilde{Y}_{n-1}]=-\tilde{X}^n(d)\tilde{X}+\tilde{Y}_n\, ,
$$
that finishes the proof of the Lemma. $\Box$

\noindent Since vector fields $X$, $\tilde{X}$ and $\tilde{Y}$ are
linearly independent, then the dimension of Lie algebra $L_x$ is at
least 3. By (\ref{DtildeYkD}), case $\tilde{Y}_1=0$ corresponds to
$\tilde{X}(d)=0$, or $d_t+d_{t_{1}}=0$ that implies $d=A(t-t_1)$,
where $A(\tau)$ is an arbitrary differentiable function of one
variable.

\noindent Assume equation (\ref{main}) admits a nontrivial
$x$-integral and $\tilde{Y}_1\ne 0$. Consider the sequence of the
vector fields $\{\tilde{Y}_1,\tilde{Y}_2, \tilde{Y}_3, \ldots\}$.
Since $L_x$ is of finite dimension, then there exists a natural
number $N$ such that
\begin{equation}\label{tildeY_N+1}
\tilde Y_{N+1}=\gamma_1 \tilde{Y}_1+\gamma_2 \tilde{Y}_2+\ldots
+\gamma_N\tilde{Y}_N, \quad N\geq 1,
\end{equation}
and $\tilde{Y}_1$, $\tilde{Y}_2$, $\ldots$, $\tilde{Y}_N$ are
linearly independent. Therefore,
$$
D\tilde{Y}_{N+1}D^{-1}=D(\gamma_1 )D\tilde{Y}_1D^{-1}+D(\gamma_2)
D\tilde{Y}_2D^{-1}+\ldots +D(\gamma_N)D\tilde{Y}_ND^{-1}, \quad
N\geq 1\, .
$$
Due to Lemma \ref{Lemma1} and (\ref{tildeY_N+1}) the last equation
can be rewritten as
$$ -\tilde{X}^{N+1}(d)\tilde{X}+\gamma_1 \tilde{Y}_1+\gamma_2 \tilde{Y}_2+\ldots
+\gamma_N\tilde{Y}_N=
$$
$$
=D(\gamma_1 )(-\tilde{X}(d)\tilde{X}+\tilde{Y}_1)+D(\gamma_2)
(-\tilde{X}^2(d)\tilde{X}+\tilde{Y}_2)+\ldots
+D(\gamma_N)(-\tilde{X}^N(d)\tilde{X}+\tilde{Y}_N)    \, .
$$
Comparing coefficients before linearly independent vector fields
$\tilde{X}$, $\tilde{Y}_1$, $\tilde{Y}_2$, $\ldots$,
$\tilde{Y}_N$, we obtain the following system of equations
$$
\begin{array}{l}
\tilde{X}^{N+1}(d)=D(\gamma_1 )\tilde{X}(d)+D(\gamma_2)
\tilde{X}^2(d)+\ldots +D(\gamma_N)\tilde{X}^N(d)\, ,\\
\gamma_1=D(\gamma_1), \quad \gamma_2=D(\gamma_2), \quad \ldots,
\quad \gamma_N=D(\gamma_N)\, .
\end{array}
$$
Since the coefficients of the vector fields $\tilde{Y}_j$ depend only on
the variables $t,t_{\pm 1}, t_{\pm 2},...$ the factors $\gamma_j$
might depend only on these variables (see Proposition above).
Hence the system of equations implies that all coefficients
$\gamma_k$, $1\leq k\leq N$, are constants, and $d=d(t, t_1)$ is a
function that satisfies the following differential equation
\begin{equation}\label{DifEqX^N}
\tilde{X}^{N+1}(d)=\gamma_1 \tilde{X}(d)+\gamma_2
\tilde{X}^2(d)+\ldots +\gamma_N\tilde{X}^N(d)\, ,
\end{equation}
where $\tilde{X}(d)=d_t+d_{t_1}$.  Using the substitution $s=t$
and $\tau=t-t_1$, equation (\ref{DifEqX^N}) can be rewritten as
\begin{equation}
\label{DifEqalpha1} \frac{\partial^{N+1}d}{\partial
s^{N+1}}=\gamma_1\frac{\partial d}{\partial
s}+\gamma_2\frac{\partial^{2}d}{\partial s^{2}}+\ldots+\gamma_N
\frac{\partial^{N}d}{\partial s^{N}}\, ,
\end{equation}
that implies that
\begin{equation}\label{formalpha1}
d(t,t_1)=\sum_k\left(\sum_{j=0}^{m_k-1}\lambda_{k,j}(t-t_1)t^j\right)\mathrm{e}^{\alpha_k
t}\, ,
\end{equation}
for some functions  $\lambda_{k,j}(t-t_1)$, where $\alpha_k$ are
roots of multiplicity $m_k$ for characteristic equation of
(\ref{DifEqalpha1}).

Let $\alpha_0=0$, $\alpha_1$, $\ldots$, $\alpha_s$ be the distinct
roots of the characteristic equation
 (\ref{DifEqX^N}). Equation  (\ref{DifEqX^N}) can
be rewritten as
\begin{equation}\label{charpolynom}
\Lambda(\tilde{X})d:=\tilde{X}^{m_0}(\tilde{X}-\alpha_1)^{m_1}(\tilde{X}-\alpha_2)^{m_2}
\ldots(\tilde{X}-\alpha_s)^{m_s}d=0\, .
\end{equation} and
$m_0+m_1+\ldots +m_s=N+1$, $m_0\geq 1$.

Initiated by the formula (\ref{tildeY}) define a map $h\rightarrow
Y_{h}$ which assigns to any function $h=h(t,t_{\pm 1},t_{\pm
2},...)$ a vector field
$$Y_h=h\frac{\partial }{\partial t_{1}}-h_{-1} \frac{\partial }{\partial
t_{-1} }+(h+h_1)\frac{\partial}{\partial
t_2}-(h_{-1}+h_{-2})\frac{\partial}{\partial t_{-2}}+...\,.$$ For
any polynomial with constant coefficients
$P(\lambda)=c_0+c_1\lambda+...+c_m\lambda^m$  we have a formula
\begin{equation}\label{map}
P(ad _{\tilde{X}})\tilde{Y}=Y_{P(\tilde{X})h},\quad\mbox{where}
\quad ad _X Y=[X,Y],
\end{equation}
which establishes an isomorphism between the linear space $V$ of
all solutions of equation (\ref{DifEqalpha1}) and the linear space
$\tilde{V}=\mathrm{span}\{\tilde Y, \tilde Y_1, ...,\tilde Y_N \}$
of the corresponding vector fields.

Represent the function (\ref{formalpha1}) as a sum
$d(t,t_1)=P(t,t_1)+Q(t,t_1)$ of the polynomial part
$P(t,t_1)=\sum_{j=0}^{m_0-1}\lambda_{0,j}(t-t_1)t^j$ and the
"exponential" part $Q(t,t_1)=\sum_{k=1}^s\left(\sum_{j=0}^{m_k-1}
\lambda_{k,j}(t-t_1)t^j\right)\mathrm{e}^{\alpha_kt}$.
\begin{lemma}
\label{Lemma23} Assume equation (\ref{main}) admits a nontrivial
$x$-integral. Then one of the functions $P(t,t_1)$ and $Q(t,t_1)$
vanishes.
\end{lemma}

\noindent{\bf Proof}. Assume in contrary that neither of the
functions vanish. First we show that in this case algebra $L_x$
contains vector fields $T_0=Y_{A(\tau)\mathrm{e}^{\alpha_k t}}$ and
$T_1=Y_{B(\tau)}$ for some functions $A(\tau)$ and $B(\tau)$.
Indeed, take $T_0:=\Lambda_0(ad_ {\tilde{X}})\tilde{Y}=
Y_{\Lambda_0(\tilde{X})d}\in L_x$, where
$\Lambda_0(\lambda)=\frac{\Lambda(\lambda)}{\lambda-\alpha_k}$.
Evidently the function $\tilde A(t,t_1)=\Lambda_0(\tilde{X})d$
solves the equation $(\tilde X -\alpha_k)\tilde
A(t,t_1)=\Lambda(\tilde{X})d=0$ which implies immediately that
$\tilde A(t,t_1)=A(\tau)\mathrm{e}^{\alpha_kt}$. In a similar way
one shows that $T_1\in L_x$. Note that due to our assumption the
functions $A(\tau)$ and $B(\tau)$ cannot vanish identically.

Consider an infinite sequence of the vector fields defined as
follows
$$ T_2=[T_0, T_1], \quad T_3=[T_0, T_2], \quad \ldots,\quad
T_n=[T_0, T_{n-1}], \quad n\geq 3.
$$
One can show that
$$
[\tilde{X}, T_0]=\alpha_k T_0, \quad [\tilde{X}, T_1]=0,\quad
[\tilde{X}, T_n]=\alpha_k(n-1) T_n,\quad n\geq 2,
$$
$$
DT_0D^{-1}=-A\mathrm{e}^{\alpha_k t}\tilde{X}+T_0, \quad
DT_1D^{-1}=-B\tilde{X}+T_1,
$$
$$
DT_nD^{-1}=T_n-\frac{(n-1)(n-2)}{2}\alpha_kA\mathrm{e}^{\alpha_k
t}T_{n-1}+b_n\tilde{X}+\sum_{k=0}^{n-2}a_{k}^{(n)}T_k, \quad n\geq
2.
$$
Since algebra $L_x$ is of finite dimension then there exists
number $N$ such that
\begin{equation}\label{111}
T_{N+1}=\lambda \tilde{X}+\mu_0T_0+\mu_1T_1+\ldots +\mu_NT_N\, ,
\end{equation}
and vector fields $\tilde{X}$, $T_0$, $T_1$,
$\ldots$, $T_N$ are linearly independent. We have,
$$
DT_{N+1}D^{-1}=D(\lambda)
\tilde{X}+D(\mu_0)\{-A\mathrm{e}^{\alpha_k
t}\tilde{X}+T_0\}+\ldots
+D(\mu_N)\Big\{T_N-\frac{(N-1)(N-2)}{2}\alpha_kA\mathrm{e}^{\alpha_k
t}T_{N-1}+\ldots\Big\}.
$$
By comparing the coefficients before $T_N$ in the last equation
one gets
$$
\mu_N-\frac{N(N-1)}{2}\alpha_kA(\tau)\mathrm{e}^{\alpha_k
t}=D(\mu_N).
$$
It follows that $\mu_N$ is a function of variable $t$ only. Also, by
applying $ad_{\tilde{X}}$ to both sides of the equation (\ref{111}),
one gets
$$
N\alpha_kT_{N+1}=[\tilde{X},
T_{N+1}]=\tilde{X}(\lambda)\tilde{X}+(\tilde{X}(\mu_0)+\mu_0\alpha_k)T_0+\ldots+
(\tilde{X}(\mu_N)+\mu_N(N-1)\alpha_k)T_N.
$$
Again, by comparing coefficients before $T_N$, we have
$$N\alpha_k\mu_N=\tilde{X}(\mu_N)+(N-1)\alpha_k\mu_N, \quad \mbox{i.e., } \quad \tilde{X}(\mu_N)=\alpha_k \mu_N. $$
Therefore, $\mu_N=A_1\mathrm{e}^{\alpha_k t}$, where $A_1$ is some
nonzero constant, and thus $A(\tau)\mathrm{e}^{\alpha_k
t}=A_2\mathrm{e}^{\alpha_k t}-A_2\mathrm{e}^{\alpha_k t_1}$. Here
$A_2$ is some constant. We have, $T_0=A_2\mathrm{e}^{\alpha_k
t}\tilde{X}-A_2S_0$, where
$$ S_0=\sum_{j=-\infty}^{\infty} \mathrm{e}^{\alpha_k
t_j}\frac{\partial}{\partial t_j}\,.
$$
Also,
$$[\tilde{X}, S_0]=\alpha_kS_0, \quad DS_0D^{-1}=S_0\, .$$
Consider a new sequence of vector fields
$$
P_1=S_0,\quad P_2=[T_1, S_0],\quad P_3=[T_1, P_2], \quad P_n=[T_1,
P_{n-1}], \quad n\geq 3\, .
$$
One can show that
$$
[\tilde{X}, P_n]=\alpha_kP_n, \quad
DP_nD^{-1}=P_n-\alpha_k(n-1)BP_{n-1}+b_n\tilde{X}+a_nS_0+\sum_{j=2}^{n-2}
a_j^{(n)}P_j, \quad n\geq 2\, .$$ Since algebra $L_x$ is of finite
dimension, then there exists number $M$ such that
\begin{equation}\label{2*}
P_{M+1}=\lambda^*\tilde{X}+\mu^*_2P_2+\ldots+\mu^*_MP_M,
\end{equation}
and fields  $\tilde{X}$, $P_2$, $\ldots$, $P_M$ are linearly
independent. Thus,
$$
DP_{M+1}D^{-1}=D(\lambda^*)\tilde{X}+D(\mu^*_2)\{P_2+\ldots\}+\ldots+D(\mu^*_M)\{P_M-\alpha_k(M-1)BP_{M-1}+\ldots\}.
$$
We compare the coefficients before $P_{M}$ in the last equation
and get
\begin{equation}\label{contradiction}\mu^*_M-M\alpha_kB(\tau)=D(\mu^*_M), \end{equation}
that implies that $\mu^*_M$ is a function of variable $t$ only.
Also,by applying $ad_{\tilde{X}}$ to both sides of (\ref{2*}), one gets
$$
\alpha_kP_{M+1}=[\tilde{X},
P_{M+1}]=\tilde{X}(\lambda^*)\tilde{X}+(\tilde{X}(\mu^*_2)+\alpha_k\mu^*_2)P_2+\ldots
+ (\tilde{X}(\mu^*_M)+\alpha_k\mu^*_M)P_M.
$$
Again, we compare the coefficients before $P_{M}$ and have
$\alpha_k\mu^*_M(t)=\tilde{X}(\mu^*_M(t))+\alpha_k\mu^*_M(t)$, that
implies that $\mu^*_M$ is a constant. It follows then from
(\ref{contradiction}) that $B(\tau)=0$. This contradiction shows
that our assumption that both functions are not identically zero
was wrong. $\Box$

\section{Multiple zero root}

In this section we assume that equation (\ref{main}) admits a
nontrivial $x$-integral and that $\alpha_0=0$ is a root of the
characteristic polynomial $\Lambda(\lambda)$. Then, due to Lemma
\ref{Lemma23}, zero is the only root and therefore
$\Lambda(\lambda)=\lambda^{m+1}$. It follows from the formula
 (\ref{formalpha1}) with $m_0=m+1$ that
$$
d(t,t_1)=a(\tau)t^m+b(\tau)t^{m-1}+\ldots,\quad m=m_0-1\geq 0.
$$
The case $m=0$ corresponds to a very simple equation $
t_{1x}=t_x+A(t-t_1), $ which is easily solved in quadratures, so
we concentrate on the case $m\geq 1$. For this case the
characteristic algebra $L_x$ contains a vector field
$T=Y_{\tilde{\kappa}}$ with
$$\tilde{\kappa}=a(\tau)t+\frac{1}{m}b(\tau).$$ Indeed,
\begin{equation}\label{ttt}
T=\frac{1}{m!}ad_{\tilde{X}}^{m-1}\tilde{Y}=Y_{\tilde{\kappa}}\,
.\end{equation}
Introduce a sequence of multiple commutators defined as follows
$$T_0=\tilde{X}, \quad T_1=[T, T_0]=Y_{-a(\tau)}, \quad T_{k+1}=[T,
T_k], \quad k\geq 0, \quad T_{k,0}=[T_0, T_k]. $$ Note that
$T_{1,0}=0$. We will see below that the linear space spanned by
this sequence is not invariant under the action of the shift
automorphism $Z\rightarrow DZD^{-1}$ introduced above. We extend
the sequence to provide the invariance property. We define
$T_{\alpha}$ with the multi-index $\alpha$.  For any sequence
$\alpha=k, 0, i_1, i_2, \ldots, i_{n-1}, i_{n}$, where $k$ is any
natural number, $i_j\in\{0;1\}$, denote
$$
T_\alpha= \left\{
\begin{array}{cl}
\left[ T_0, T_{k, 0,i_1, \ldots, i_{n-1}} \right], & {\rm{if}}\quad i_n=0;\\
\\
\left[ T, T_{k, 0,i_1, \ldots, i_{n-1}} \right],& {\rm{if}} \quad i_n=1;\\
\end{array}
\right.
$$
$$
m(\alpha)=\left\{\begin{array}{cl}
k, &{\rm{if}} \quad\alpha=k;\\
k,& {\rm{if}}\quad \alpha=k,0;\\
k+i_1+\ldots+i_n,& {\rm{if}} \quad
\alpha=k,0,i_1,\ldots,i_n;\end{array}\right.
$$
$$
l(\alpha)=k+n+1-m(\alpha).
$$
The multi-index $\alpha$ is characterized by two quantities
$m(\alpha)$ and $l(\alpha)$ which allow to order partially the
sequence $\{T_\alpha\}$. We have,
$$
DT_0D^{-1}=T_0, \quad DTD^{-1}=T-\tilde{\kappa} T_0, \quad
DT_1D^{-1}=T_1+aT_0.$$ One can prove by induction on $k$ that
\begin{equation}\label{e2}
DT_kD^{-1}=T_k+aT_{k-1}-\tilde{\kappa}\sum_{m(\beta)=k-1
}T_\beta+\sum_{m(\beta)\leq k-2}\eta(k,\beta)T_\beta\,.
\end{equation}
In general, for any $\alpha$,
\begin{equation}
\label{e22} DT_\alpha D^{-1}=T_\alpha+\sum_{m(\beta)\leq
m(\alpha)-1}\eta(\alpha, \beta)T_\beta\, . \end{equation}
 We can choose a system $P$ of linearly independent vector fields in the following way.\\
1) $T$ and $T_0$ are linearly independent. We take them into $P$. \\
2) We check whether $T$, $T_0$ and $T_1$ are linearly independent
or not. If they are dependent then $P=\{T, T_0\}$ and
$T_1=\mu
T+\lambda T_0$ for some functions $\mu$ and $\lambda$.\\
3) If $T$, $T_0$, $T_1$ are linearly independent then we check
whether $T$, $T_0$, $T_1$, $T_2$ are linearly independent or not.
If
they are dependent, then  $P=\{T, T_0, T_1\}$. \\
4) If $T$, $T_0$, $T_1$, $T_2$ are linearly independent, we add
vector fields $T_\beta$, $m(\beta)=2$, $\beta\in I_2$, (actually,
by definition $I_2$ is the collection of such $\beta$) in such a
way that $J_2:=\{T, T_0, T_1, T_2, \cup_{\beta\in I_2} T_\beta\}$
is a system of linearly independent vector fields and for any
$T_\gamma$ with $m(\gamma)\leq 2$ we have
$T_\gamma=\sum\limits_{T_\beta\in
J_2}\mu(\gamma, \beta)T_\beta$.\\
5) We check whether $T_3\cup J_2$ is a linearly independent system.
If it is not, then  $P$ consists of all elements from $J_2$,
and $T_3=\sum\limits_{T_\beta\in J_2}\mu(\gamma, \beta)T_\beta$. If
it is, then to the system $T_3\cup J_2$ we add vector fields
$T_\beta$, $m(\beta)=3$, $\beta\in I_3$, in such a way that
$J_3:=\{T_3, J_2, \cup_{\beta\in I_3}T_\beta\}$ is a system of
linearly independent vector fields and for any $T_\gamma$ with
$m(\gamma)\leq 3$ we have
 $T_\gamma=\sum\limits_{T_\beta\in
J_3}\mu(\gamma, \beta)T_\beta$.\\
We continue the construction of the system  $P$. Since $L_x$ is of
finite
dimension, then there exists such a natural number $N$ that \\
(i) $T_k\in P$, $k\leq N$; \\
(ii) $m(\beta)\leq N$ for any $T_\beta\in P$; \\
(iii) for any $T_\gamma$ with $m(\gamma)\leq N$ we have
$T_\gamma=\sum\limits_{T_\beta\in P, m(\beta)\leq
m(\gamma)}\mu(\gamma, \beta)T_\beta$ and also $$T_{N+1}=\mu(N+1,
N)T_N+\sum\limits_{T_\beta\in
P,m(\beta)\leq N }\mu(N+1, \beta)T_\beta.$$ \\
It follows that\\
(iv) for any vector field $T_\alpha$ with $m(\alpha)=N$,  that does
not belong to $P$, the coefficient $\mu(\alpha, N)$ before $T_N$ in
the expansion
\begin{equation}\label{asli34}
T_\alpha=\mu(\alpha, N)T_N+\sum\limits_{T_\beta\in P}\mu(\alpha,
\beta)T_\beta
\end{equation}
is constant. Indeed, by (\ref{e22}),
$$
DT_\alpha D^{-1}=T_\alpha+\sum\limits_{m(\beta)\leq
N-1}\eta(\alpha,\beta)T_\beta=\mu(\alpha,
N)T_N+\sum\limits_{T_\beta\in P}\mu(\alpha,
\beta)T_\beta+\sum\limits_{m(\beta)\leq
N-1}\eta(\alpha,\beta)T_\beta\, .
$$
From (\ref{asli34}) we have also
\begin{eqnarray*}
DT_\alpha D^{-1}&=&D(\mu(\alpha,
N))DT_ND^{-1}+\sum\limits_{T_\beta\in P}D(\mu(\alpha,
\beta))DT_\beta D^{-1}\\
&=&D(\mu(\alpha, N))\{T_N+\ldots\}+\sum\limits_{T_\beta\in
P}D(\mu(\alpha, \beta))\{T_\beta+\ldots\}\, .
\end{eqnarray*}
\noindent By comparing the coefficients before $T_N$ in these two
expressions for $DT_\alpha D^{-1}$, we have
$$
\mu(\alpha, N)=D(\mu(\alpha, N)),
$$ that implies that $\mu(\alpha, N)$ is a constant indeed.

\begin{lemma}\label{LemmaZero1}
We have, $a(\tau)=c_0\tau+c_1$, where $c_0$ and $c_1$ are some
constants.
\end{lemma}

\noindent{\bf Proof}. Since
$$T_{N+1}=\mu(N+1, N)T_N+\sum\limits_{T_\beta\in P}\mu(N+1,
\beta)T_\beta\, ,$$ then
$$
DT_{N+1}D^{-1}=D(\mu(N+1, N))\{T_N+\ldots\}+\sum\limits_{T_\beta\in
P}D(\mu(N+1, \beta))\{T_\beta+\ldots\}.
$$
On the other hand,
$$
DT_{N+1}D^{-1}=T_{N+1}+aT_N- \tilde{\kappa}\sum_{m(\beta)=N
}T_\beta+\sum_{m(\beta)\leq
N-1}\eta(N+1,\beta)T_\beta\, .
$$
We compare the coefficients before $T_N$ in the last two
expressions. For $N\geq 0$ the equation is
\begin{equation}\label{e3}
\mu(N+1, N)+a-\tilde{\kappa}\sum\limits_{T_\beta\in P, m(\beta)=N}\mu(\beta, N)=D(\mu(N+1, N))\, .
\end{equation}
Denote by $c=-\sum\limits_{T_\beta\in P, m(\beta)=N}\mu(\beta, N)$
and by $\mu_N=\mu(N+1, N)$. By property (iv), $c$ is a constant.
It follows from (\ref{e3}) that $\mu_N$ is a function of variables
$t$ and $n$ only. Therefore,
$$
a(\tau)+c\left(a(\tau)t+\frac{1}{m}b(\tau)\right)=\mu_N(t_1, n+1)-\mu_N(t,
n).
$$
By differentiating both sides of the equation with respect to $t$
and then $t_1$, we have
$$-a''(\tau)-c\left( a''(\tau)t+a'(\tau)+\frac{1}{m}b''(\tau)\right)=0,
$$ that implies that $a''(\tau)=0$, or the same, $a(\tau)=c_0\tau
+c_1$ for some constants $c_0$ and $c_1$. $\Box$

Vector fields $T_1$ and $T$ in new variables are rewritten as
\begin{equation}\label{nata1}
T_1=\sum\limits_{j=-\infty}^\infty
a(\tau_j)\frac{\partial}{\partial \tau_j}\, ,
\end{equation}
\begin{eqnarray}
T=-\sum\limits_{j=-\infty}^\infty
\{a(\tau_j)t_j+\frac{1}{m}b(\tau_j)\}\frac{\partial}{\partial
\tau_j} &=&-\sum\limits_{j=-\infty}^\infty
\{a(\tau_j)(t+\rho_j)+\frac{1}{m}b(\tau_j)\}\frac{\partial}{\partial
\tau_j}\nonumber\\&=& -tT_1-\sum\limits_{j=-\infty}^\infty
\{a(\tau_j)\rho_j+\frac{1}{m}b(\tau_j)\}\frac{\partial}{\partial
\tau_j}\, ,\label{nata2}
\end{eqnarray}
where
$$
\rho_j=\left\{\begin{array}{cl} -\tau-\tau_1-\ldots-\tau_{j-1}, &
{\rm{if}}
\quad j\geq 1;\\
0, & {\rm{if}}
\quad j=0;\\
\tau_{-1}+\tau_{-2}+\ldots+\tau_{j}, & {\rm{if}} \quad j\leq -1\,
.
\end{array}
\right.
$$
The following two lemmas are to be useful.

\begin{lemma}\label{LemmaAux1}
If the Lie algebra generated by the vector fields
$S_0=\sum\limits_{j=-\infty}^\infty \frac{\partial}{\partial w_j}$
and $P=\sum\limits_{j=-\infty}^\infty
c(w_j)\frac{\partial}{\partial
w_j}$ is of finite dimension then $c(w)$ is one of the forms\\
(1) $c(w)=c_2+c_3\mathrm{e}^{\lambda w}+c_4\mathrm{e}^{-\lambda w}$, $\lambda\ne 0$;\\
(2) $c(w)=c_2+c_3w+c_4w^2$, where $c_2$, $c_3$, $c_4$ are some
constants.
\end{lemma}
\noindent \textbf{Proof.} Introduce vector fields
$$S_1=[S_0,P],\quad S_2=[S_0,S_1],\quad ...,\quad S_n=[S_0,S_{n-1}],\quad n\geq 3.$$
Clearly, we have
\begin{equation}\label{multiplezeroS_n}
S_n=\sum_{j=-\infty}^{\infty}c^{(n)}(w_j)\frac{\partial}{\partial
w_j},\quad n\geq 1.
\end{equation}
Since all vector fields $S_n$ are elements of $L_x$, and $L_x$ is
of finite dimension, then there exists a natural number $N$ such
that
\begin{equation}\label{multiplezeroS_(N+1)}
S_{N+1}=\mu_NS_N+\mu_{N-1}S_{N-1}+...+\mu_1S_1+\mu_0P+\mu S_0,
\end{equation}
and $S_0,P,S_1,...,S_N$ are linearly independent. (Note that we
may assume $S_0$ and $P$ are linearly independent). Since
$DS_0D^{-1}=S_0$, $DPD^{-1}=P$ and $DS_nD^{-1}=S_n$ for any $n\geq
1$, then it follows from (\ref{multiplezeroS_(N+1)}) that
$$S_{N+1}=D(\mu_N)S_N+D(\mu_{N-1})S_{N-1}+...+D(\mu_1)S_1+D(\mu_0)P+D(\mu)S_0$$
and together with (\ref{multiplezeroS_(N+1)}), it implies that
$\mu, \mu_0, \mu_1, ..., \mu_N$ are all constants.

\noindent By comparing the coefficients before
$\frac{\partial}{\partial w}$ in (\ref{multiplezeroS_(N+1)}) one
gets, with the help of (\ref{multiplezeroS_n}), the following
equality
$$c^{(N+1)}(w)=\mu_Nc^{(N)}(w)+...+\mu_1c'(w)+\mu_0c(w)+\mu.$$
Thus, $c(w)$ is a solution of the nonhomogeneous linear
differential equation with constant coefficient whose
characteristic polynomial is
$$\Lambda(\lambda)=\lambda^{N+1}-\mu_N\lambda^{N}-...-\mu_1\lambda-\mu_0.$$
Denote by $\beta_1, \beta_2, ..., \beta_t$ characteristic roots
and by $m_1, m_2, ..., m_t$ their multiplicities. There are the
following possibilities:
\begin{enumerate}
\item[(i)] There exists a nonzero characteristic root, say
$\beta_1$, and its multiplicity $m_1\geq 2$,\\
\item[(ii)] There exists zero characteristic root, say $\beta_1$,
and $m_1\geq 3$, $\mu=0$ or $m_1\geq 2$, $\mu\neq 0$,\\
\item[(iii)] There are two distinct characteristic roots, say
$\beta_1$ and $\beta_2$ with $\beta_1\neq 0$, $\beta_2=0$,\\
\item[(iv)] There are two nonzero distinct characteristic roots,
say $\beta_1$ and $\beta_2$.
\end{enumerate}
\noindent In case $\rm{(i)}$, consider
$$\Lambda_1(\lambda)=\frac{\Lambda(\lambda)}{\lambda-\beta_1}
\quad \rm{and} \quad
\Lambda_1^{(2)}(\lambda)=\frac{\Lambda(\lambda)}{(\lambda-\beta_1)^2}.$$
Then $\Lambda_1(S_0)c(w)=\alpha_1\mathrm{e}^{\beta_1w}+\alpha_2$ and
$\Lambda_1^{(2)}(S_0)c(w)=(\alpha_3w+\alpha_4)\mathrm{e}^{\beta_1w}+\alpha_5$,
where $\alpha_j$, $1\leq j \leq 5$, are some constants with
$\alpha_1\neq 0$, $\alpha_3\neq 0$. We have,
\begin{eqnarray*}
\Lambda_1(ad_{S_0})P&=&\sum_{j=-\infty}^{\infty}(\alpha_1\mathrm{e}^{\beta_1w_j}+\alpha_2)\frac{\partial}{\partial
w_j}=\alpha_1\Big(\sum_{j=-\infty}^{\infty}\mathrm{e}^{\beta_1w_j}\frac{\partial}{\partial
w_j}\Big)+\alpha_2S_0=\alpha_1P_1+\alpha_2S_0,\\\\
\Lambda_1^{(2)}(ad_{S_0})P&=&\sum_{j=-\infty}^{\infty}((\alpha_3w_j+\alpha_4)e^{\beta_1
w_j}+\alpha_5)\frac{\partial}{\partial
w_j}=\alpha_3\Big(\sum_{j=-\infty}^{\infty}w_j\mathrm{e}^{\beta_1w_j}\frac{\partial}{\partial
w_j}\Big)+\alpha_4P_1+\alpha_5S_0\\
&=&\alpha_3P_2+\alpha_4P_1+\alpha_5S_0
\end{eqnarray*}
are elements from $L_x$ and therefore vector fields
$P_1=\sum_{j=-\infty}^{\infty}\mathrm{e}^{\beta_1w_j}\frac{\partial}{\partial
w_j}$ and
$P_2=\sum_{j=-\infty}^{\infty}w_j\mathrm{e}^{\beta_1w_j}\frac{\partial}{\partial
w_j}$ belong to $L_x$. Since $P_1$ and $P_2$ generate an infinite
dimensional Lie algebra $L_x$ then case $\rm{(i)}$ fails to be
true.\\

\noindent In case $\rm{(ii)}$, consider
$$\Lambda_1^{(3)}(\lambda)=\frac{\Lambda(\lambda)}{\lambda^3}\quad
{\rm{and}}\quad
\Lambda_1^{(2)}(\lambda)=\frac{\Lambda(\lambda)}{\lambda^2}, \quad
{\rm{if}} \quad \mu=0,$$ \noindent or
$$\Lambda_1^{(3)}(\lambda)=\frac{\Lambda(\lambda)}{\lambda^2}\quad
{\rm{and}}\quad
\Lambda_1^{(2)}(\lambda)=\frac{\Lambda(\lambda)}{\lambda}, \quad
{\rm{if}}\quad \mu\neq 0.$$

\noindent We have
$$\Lambda_1^{(3)}(S_0)c(w)=\alpha_1w^3+\alpha_2w^2+\alpha_3w+\alpha_4\quad {\rm{and}} \quad
\Lambda_1^{(2)}(S_0)c(w)=\alpha_5w^2+\alpha_6w+\alpha_7,$$

\noindent where $\alpha_j$, $1\leq j\leq 7$, are some constants
with $\alpha_1\neq 0$, $\alpha_5\neq 0$. Straightforward
calculations show that vector fields
$$\Lambda_1^{(3)}(ad_{S_0})P
=\sum_{j=-\infty}^{\infty}(\alpha_1w_j^3+\alpha_2w_j^2
+\alpha_3w_j+\alpha_4)\frac{\partial}{\partial w_j}\quad {\rm{and}}
\quad
\Lambda_1^{(2)}(ad_{S_0})P=\sum_{j=-\infty}^{\infty}(\alpha_5w_j^2+\alpha_6w_j+\alpha_7)\frac{\partial}{\partial
w_j}
$$
generate an infinite dimensional Lie algebra. It proves that case
$\rm{(ii)}$ fails to be true.\\

\noindent In case $\rm{(iii)}$, consider
$$\Lambda_1(\lambda)=\frac{\Lambda(\lambda)}{\lambda-\beta_1}\quad \rm{and}\quad
\Lambda_2(\lambda)=\frac{\Lambda(\lambda)}{\lambda}.$$

\noindent We have
$$\Lambda_1c(w)=\alpha_1\mathrm{e}^{\beta_1w}+\alpha_2\quad {\rm{and}}\quad
\Lambda_2c(w)=\alpha_3w+\alpha_4,\quad \rm{if}\quad \mu=0,$$

\noindent or

$$\Lambda_1(S_0)c(w)=\alpha_1\mathrm{e}^{\beta_1w}+\alpha_2\quad {\rm{and}}\quad
\Lambda_2(S_0)c(w)=\alpha_5w^2+\alpha_6w+\alpha_7,\quad \rm{if}\quad
\mu\neq 0,$$

\noindent where $\alpha_j$, $1\leq j\leq 7,$ are constants with
$\alpha_1\neq 0$, $\alpha_3\neq 0$, $\alpha_5\neq 0$. Since vector
fields $\Lambda_1(ad_{S_0})P$ and $\Lambda_2(ad_{S_0})P$ generate
an infinite dimensional Lie algebra, then case $\rm{(iii)}$ also
fails to
exist.\\

\noindent In case $\rm{(iv)}$, consider
$$\Lambda_1(\lambda)=\frac{\Lambda(\lambda)}{\lambda-\beta_1}\quad
{\rm{and}}
\quad
\Lambda_2(\lambda)=\frac{\Lambda(\lambda)}{\lambda-\beta_2}.$$
We have,
$\Lambda_1(S_0)c(w)=\alpha_1\mathrm{e}^{\beta_1w}+\alpha_2$,
$\Lambda_2(S_0)c(w)=\alpha_3\mathrm{e}^{\beta_2w}+\alpha_4$, where
$\alpha_1\neq 0$, $\alpha_2$, $\alpha_3\neq 0$, $\alpha_4$ are some
constants. Note that
$$\Lambda_1(ad_{S_0})P=\alpha_1\Big(\sum_{j=-\infty}^{\infty}\mathrm{e}^{\beta_1w_j}\frac{\partial}{\partial w_j}\Big)+\alpha_2S_0
\quad {\rm{and}}\quad
\Lambda_2(ad_{S_0})P=\alpha_3\Big(\sum_{j=-\infty}^{\infty}\mathrm{e}^{\beta_2w_j}\frac{\partial}{\partial
w_j}\Big)+\alpha_4S_0,$$

\noindent and vector fields
$\sum_{j=-\infty}^{\infty}\mathrm{e}^{\beta_1w_j}\frac{\partial}{\partial
w_j}$ and
$\sum_{j=-\infty}^{\infty}\mathrm{e}^{\beta_2w_j}\frac{\partial}{\partial
w_j}$ generate an infinite dimensional Lie algebra if
$\beta_1+\beta_2\neq 0$.

\noindent It follows from $\rm{(i)}$, $\rm{(ii)}$, $\rm{(iii)}$,
$\rm{(iv)}$
that $c(w)$ is one of the forms\\
(1) $c(w)=c_2+c_3\mathrm{e}^{\lambda w}+c_4\mathrm{e}^{-\lambda w}$, $\lambda\ne 0$;\\
(2) $c(w)=c_2+c_3w+c_4w^2$, where $c_2$, $c_3$, $c_4$ are some
constants. $\Box$

\begin{lemma}\label{LemmaAux2}
If the Lie algebra generated by the vector fields
$S_0=\sum\limits_{j=-\infty}^\infty \frac{\partial}{\partial
w_j}$, $Q=\sum\limits_{j=-\infty}^\infty
q(w_j)\frac{\partial}{\partial w_j}$  and
$S_1=\sum\limits_{j=-\infty}^\infty
\{\tilde{\rho}_j+\tilde{b}(w_j)\}\frac{\partial}{\partial w_j}$ is
of finite dimension then $q(w)$ is a constant function.
\end{lemma}

\noindent \textbf{Proof.} It follows from Lemma \ref{LemmaAux1}
that
\begin{enumerate}
\item[(1)] $q(w)=c_2+c_3w+c_4w^2$, or\\
\item[(2)] $q(w)=c_2+c_3\mathrm{e}^{\lambda
w}+c_4\mathrm{e}^{-\lambda w}$, $\lambda\neq 0$,
\end{enumerate}
\noindent where $c_2$, $c_3$, $c_4$ are some constants.

\noindent Consider case $(1)$. We have,
$$[S_0,Q]=c_3\sum_{j=-\infty}^{\infty}\frac{\partial}{\partial w_j}
+2c_4\sum_{j=-\infty}^{\infty}w_j\frac{\partial}{\partial w_j}
=c_3S_0+2c_4\sum_{j=-\infty}^{\infty}w_j\frac{\partial}{\partial
w_j}.$$

\noindent If $c_4\neq 0$, then
$\sum_{j=-\infty}^{\infty}w_j\frac{\partial}{\partial w_j} \in
L_x$ and $\sum_{j=-\infty}^{\infty}w_j^2\frac{\partial}{\partial
w_j} \in L_x$.\\

\noindent If $c_4=0$, $c_3\neq 0$, then
$\sum_{j=-\infty}^{\infty}w_j\frac{\partial}{\partial
w_j}=\frac{1}{c_3}(Q-c_2S_0) \in L_x$.\\

\noindent If $c_3=c_4=0$, then $q(w)=c_2$ and there is nothing to prove.\\

\noindent Assume $c_4^2+c_3^2\neq 0$. Denote by
$P=\sum_{j=-\infty}^{\infty}w_j\frac{\partial}{\partial w_j}$.
Construct the vector fields
$$P_1=[P,S_1],\,P_n=[P,P_{n-1}],\quad n\geq 2.$$
\noindent We have,
\begin{eqnarray*}
DS_0D^{-1}&=&S_0,\\
DS_1D^{-1}&=&S_1-(\mathrm{e}^w-\tilde{c})S_0,\\
DPD^{-1}&=&P,\\
DP_1D^{-1}&=&P_1+(-w\mathrm{e}^w+\mathrm{e}^w-\tilde{c})S_0,\\
DP_2D^{-1}&=&P_2+(-w^2\mathrm{e}^w+w\mathrm{e}^w-\mathrm{e}^w+\tilde{c})S_0.
\end{eqnarray*}
\noindent In general,
$$DP_nD^{-1}=P_n+(-w^n\mathrm{e}^w+R_{n-1}(w)\mathrm{e}^w+c_n)S_0,\quad n\geq 3,$$
where $R_{n-1}$ is a polynomial of degree $n-1$, and $c_n$ is a
constant. Since $L_x$ is of finite dimension, then there exists a
natural number $N$ such that
$$P_{N+1}=\mu_NP_N+...+\mu_1P_1+\mu_0S_0,$$
and $S_0, P_1,..., P_N$ are linearly independent. Thus
$$DP_{N+1}D^{-1}=D(\mu_N)DP_ND^{-1}+...+D(\mu_1)DP_1D^{-1}+D(\mu_0)S_0,$$
\noindent or the same,
\begin{eqnarray*}
&&\mu_NP_N+...+\mu_1P_1+\mu_0S_0+(-w^{N+1}\mathrm{e}^w+R_N(w)\mathrm{e}^w+c_{N+1})S_0\\
&&=D(\mu_N)\{P_N+(-w^N\mathrm{e}^w+R_{N-1}(w)\mathrm{e}^w+c_N)S_0\}+...\\
&&+D(\mu_1)\{P_1+(-w\mathrm{e}^w+\mathrm{e}^w-\tilde{c})S_0\}+D(\mu_0)S_0.
\end{eqnarray*}
\noindent By comparing the coefficients before $P_N,...,P_1$ we
have
$$\mu_N=D(\mu_N),\,...\,,\,\mu_1=D(\mu_1),$$
that implies $\mu_N,...,\mu_1$ are all constants. By comparing the
coefficients before $S_0$ we have \begin{eqnarray*}
\mu_0-w^{N+1}\mathrm{e}^w+R_N(w)\mathrm{e}^w+c_{N+1}&=&\mu_N(-w^N\mathrm{e}^w+R_{N-1}(w)\mathrm{e}^w+c_N)\\
&&+... +\mu_1(-w\mathrm{e}^w+\mathrm{e}^w-\tilde{c})+D(\mu_0).
\end{eqnarray*}
\noindent The last equality shows that $D(\mu_0)-\mu_0$ is a
function of $w$ only. Thus $D(\mu_0)-\mu_0$ is a constant, denote it
by $d_0$. The last equality becomes a contradictory one:
\begin{eqnarray*}
w^{N+1}\mathrm{e}^w&=&R_N(w)\mathrm{e}^w+c_{N+1}-\mu_N(-w^N\mathrm{e}^w+R_{N-1}(w)\mathrm{e}^w+c_N)\\
&&-... -\mu_1(-w\mathrm{e}^w+\mathrm{e}^w-\tilde{c})-d_0.
\end{eqnarray*}
\noindent This contradiction proves that $c_3^2+c_4^2=0$, i.e.
$c_3=c_4=0$ in case $(1)$. Therefore, $q(w)=c_2$.

\noindent Consider case $(2)$. Since
$$[S_0,Q]=\lambda c_3
\sum_{j=-\infty}^{\infty}\mathrm{e}^{\lambda
w_j}\frac{\partial}{\partial w_j} -\lambda c_4
\sum_{j=-\infty}^{\infty}\mathrm{e}^{-\lambda
w_j}\frac{\partial}{\partial w_j},$$
$$[S_0,[S_0,Q]]=\lambda^2 c_3
\sum_{j=-\infty}^{\infty}\mathrm{e}^{\lambda
w_j}\frac{\partial}{\partial w_j} +\lambda^2 c_4
\sum_{j=-\infty}^{\infty}\mathrm{e}^{-\lambda
w_j}\frac{\partial}{\partial w_j},$$

\noindent then vector fields
$Q_{\lambda}=c_3\sum_{j=-\infty}^{\infty}\mathrm{e}^{\lambda
w_j}\frac{\partial}{\partial w_j}$ and
$Q_{-\lambda}=c_4\sum_{j=-\infty}^{\infty}\mathrm{e}^{-\lambda
w_j}\frac{\partial}{\partial w_j}$ both belong to $L_x$. We have,
$DQ_{\lambda}D^{-1}=Q_{\lambda}$,
$DQ_{-\lambda}D^{-1}=Q_{-\lambda}$.

\noindent Assume $c_3\neq 0$. Construct vector fields
$$Q_1=[Q_{\lambda},S_1],\quad Q_n=[Q_{\lambda},Q_{n-1}],\quad n\geq 2.$$
\noindent Direct calculations show that
\begin{eqnarray*}
DQ_1D^{-1}&=&Q_1-c_3\mathrm{e}^{(1+\lambda)w}S_0+(\mathrm{e}^w-\tilde{c})\lambda
Q_{\lambda},\\
DQ_2D^{-1}&=&Q_2-c_3^2(1+\lambda)\mathrm{e}^{(1+2\lambda)w}S_0+2\lambda
c_3\mathrm{e}^{(1+\lambda)w}Q_{\lambda}.
\end{eqnarray*}
\noindent It can be proved by induction on $n$ that
$$DQ_nQ^{-1}=Q_n-p_nS_0+q_nQ_{\lambda},\quad n\geq 2,$$
\noindent where
\begin{eqnarray*}
p_n&=&c_3^n(1+\lambda)(1+2\lambda)...(1+(n-1)\lambda)\mathrm{e}^{(1+n\lambda)w},\\
q_n&=&nc_3^{n-1}\lambda(1+\lambda)...(1+(n-2)\lambda)\mathrm{e}^{(1+(n-1)\lambda)w}.
\end{eqnarray*}
\noindent Since $L_x$ is of finite dimension, there exists such a
natural number $N$ that
$$Q_{N+1}=\mu_NQ_N+...+\mu_1Q_1+\mu_{\lambda}Q_{\lambda}+\mu_0S_0,$$
and $S_0,Q_{\lambda},Q_1,...,Q_N$ are linearly independent. Then
$$DQ_{N+1}D^{-1}=D(\mu_N)DQ_ND^{-1}+...+D(\mu_0)DS_0D^{-1},$$
\noindent or
\begin{eqnarray*}
&&\mu_NQ_N+...+\mu_1Q_1+\mu_{\lambda}Q_{\lambda}+\mu_0S_0-p_{N+1}S_0+q_{N+1}Q_{\lambda}\\
&&=D(\mu_N)\{Q_N-p_NS_0+q_NQ_{\lambda}\}+...
+D(\mu_1)\{Q_1-p_1S_0+q_1Q_{\lambda}\}\\
&&+D(\mu_{\lambda})Q_{\lambda}+D(\mu_0)S_0.
\end{eqnarray*}
\noindent By comparing the coefficients before $Q_N,...,Q_1$, we
have that $\mu_k$, $1\leq k\leq N$, are all constants. Comparing
coefficients before $S_0$ gives
\begin{equation}\label{coeffS_0}
\mu_0-p_{N+1}=-\mu_Np_N-...-\mu_2p_2-\mu_1p_1+D(\mu_0).
\end{equation}
\noindent Since $p_k$, $1\leq k \leq N+1$, depend on $w$ only,
then $D(\mu_0)-\mu_0$ is a function of $w$, and therefore
$D(\mu_0)-\mu_0$ is a constant, denote it by $d_0$.

\noindent If $\lambda\neq -\frac{1}{r}$ for all $r\in \mathbb{N}$,
then $p_k\neq 0$ for all $k\in \mathbb{N}$, and equation
(\ref{coeffS_0}) fails to be true.

\noindent Consider case when $\lambda=-\frac{1}{r}$ for some $r\in
\mathbb{N}$. Substitution $u_j=\mathrm{e}^{-\lambda w_j}$
transforms vector fields $\frac{-1}{\lambda c_3}Q_{\lambda}$,
$\frac{-1}{\lambda}S_1$, $\frac{-1}{\lambda}S_0$ into vector
fields
\begin{eqnarray*}
Q_{\lambda}^*&=&\sum_{j=-\infty}^{\infty}\frac{\partial}{\partial
u_j},\\
S_1^*&=&\sum_{j=-\infty}^{\infty}\{\tilde{\rho}_j^*+\tilde{b}^*(u_j)\}u_j\frac{\partial}{\partial
u_j},\\
S_0^*&=&\sum_{j=-\infty}^{\infty}u_j\frac{\partial}{\partial u_j},
\end{eqnarray*}
\noindent where
$$
\tilde{\rho}_j^*=\left\{\begin{array}{cl}\sum\limits_{k=0}^{j-1}(u_k^r-\tilde{c}),
& {\rm{if}}
\quad j\geq 1;\\
0, &{\rm{if}}
\quad j=0;\\
-\sum\limits_{k=j}^{-1}(u_k^r-\tilde{c}), & {\rm{if}} \quad j\leq
-1,
\end{array}
\right. , \quad \tilde{b}^*(u_j)=\tilde{b}(r\ln u_j)\, .
$$
\noindent First consider the case $r=1$. We have,
\begin{eqnarray*}
T:&=&[Q_{\lambda}^*,S_1^*]=\sum_{j=-\infty}^{\infty}\{ju_j+\tilde{\rho}_j^*+
\tilde{b}^*(u_j)+u_j{\tilde{b}^{*'}}(u_j)
\}\frac{\partial}{\partial u_j},\\
K:&=&\frac{1}{2}[Q_{\lambda}^*,T]=\sum_{j=-\infty}^{\infty}\{j+c(u_j)\}
\frac{\partial}{\partial u_j},
\end{eqnarray*}
\noindent where
$c(u_j)={\tilde{b}^{*'}}(u_j)+\frac{1}{2}u_j{\tilde{b}^{*''}}(u_j)$,
\begin{eqnarray*}
T_1&=&[T,K]=\gamma_1\sum_{j=-\infty}^{\infty}\{j^2+jg_{1,1}^{(j)}(u_j)
+g_{1,0}^{(j)}(u,u_1,...,u_j)\}\frac{\partial}{\partial u_j},\\
T_2&=&[T,T_1]=\gamma_2\sum_{j=-\infty}^{\infty}\{j^3+j^2g_{2,2}^{(j)}(u_j)+jg_{2,1}^{(j)}(u,u_1,...,u_j)
+g_{2,0}^{(j)}(u,u_1,...,u_j)\}\frac{\partial}{\partial u_j},
\end{eqnarray*}
\noindent where $\gamma_1=-\frac{3}{2}$ and $\gamma_2\neq 0$.

\noindent Construct vector fields, $T_n=[T, T_{n-1}]$, $n\geq 3$.
Direct calculations show that
$$T_n=\gamma_n\sum_{j=0}^{\infty}\Big\{j^{n+1}+j^ng_{n,n}(u_j)
+\sum_{k=0}^{n-1}j^kg_{n,k}(u,u_1,...,u_j)\Big\}\frac{\partial}{\partial
u_j} +\sum_{j=-\infty}^{-1}a_j\frac{\partial}{\partial u_j},\quad
n\geq 1.$$

\noindent Since $\{T_n \}_{n=1}^{\infty}$ is an infinite sequence
of linearly independent vector fields from $L_x$, then case $r=1$
fails to exist.

\noindent Consider case $r\geq 2$. We have,
$$ad_{Q_{\lambda}^*}S_1^*=[Q_{\lambda}^*,S_1^*]=\sum_{j=-\infty}^{\infty}
\Big\{sgn(j)r\Big(\sum_{k=0}^{j-1}u_k^{r-1}\Big)u_j+\tilde{\rho}_j^*+\tilde{b}^*(u_j)
+u_j\tilde{b}^{*'}(u_j)\Big \}\frac{\partial}{\partial u_j},$$ and
$$ad_{Q_{\lambda}^*}^rS_1^*=\sum_{j=-\infty}^{\infty}\Big\{r!ju_j+sgn(j)r!\sum_{k=0}^{j-1}u_k+d(u_j)\Big\}$$
for some function $d$,
$$ad_{Q_{\lambda}^*}^{r+1}S_1^*=\sum_{j=-\infty}^{\infty}\Big\{2r!j+d'(u_j)\Big\}\frac{\partial}{\partial u_j}.$$
Note that vector fields $ad_{Q_{\lambda}^*}^rS_1^*$ and
$ad_{Q_{\lambda}^*}^{r+1}S_1^*$ have coefficients of the same kind
as vector fields $T$ and $K$ (from case $r=1$) have. It means that
$ad_{Q_{\lambda}^*}^rS_1^*$ and $ad_{Q_{\lambda}^*}^{r+1}S_1^*$
generate an infinite dimensional Lie algebra. This contradiction
implies that case $r\geq 2$ also fails to exist.

\noindent Thus, $c_3=0$. By interchanging $\lambda$ with
$-\lambda$, we obtain that $c_4=0$ also. Hence $c_3=c_4=0$ and
$q(w)=c_2$. $\Box$

We already know that $a(\tau)=c_0\tau+c_1$. The next lemma shows that
$c_0\ne 0$.


\begin{lemma}\label{LemmaZero2}
$c_0$ is a nonzero constant.
\end{lemma}

\noindent{\bf Proof}. Assume contrary. Then $a(\tau)=c_1$ and
$c_1\ne 0$, vector fields (\ref{nata1}) and (\ref{nata2}) become
$$
T_1=c_1\sum\limits_{j=-\infty}^\infty \frac{\partial}{\partial
\tau_j}=c_1\tilde{T}_1,$$ and $$
T=-tT_1-c_1\sum\limits_{j=-\infty}^\infty
\{\rho_j+\frac{1}{mc_1}b(\tau_j)\}\frac{\partial}{\partial \tau_j}\,
=-c_1t\tilde{T}_1-c_1\tilde{T},
$$
where $$ \tilde{T}_1=\sum\limits_{j=-\infty}^\infty
\frac{\partial}{\partial \tau_j}, \qquad
\tilde{T}=\sum\limits_{j=-\infty}^\infty
\{\rho_j+\frac{1}{mc_1}b(\tau_j)\}\frac{\partial}{\partial
\tau_j}.
$$ Since
$$
[\tilde{T_1},[\tilde{T_1},\tilde{T}]]=\frac{1}{mc_1}\sum_{j=-\infty}^\infty
b''(\tau_j)\frac{\partial }{\partial \tau_j}
$$
and $\tilde{T}_1$ both belong to a finite dimensional $L_x$, then, by Lemma \ref{LemmaAux1},  1)
$b''(\tau)=\tilde{C}_1+\tilde{C}_2\mathrm{e}^{\lambda
\tau}+\tilde{C}_3\mathrm{e}^{-\lambda\tau}$ or 2)
$b''(\tau)=\tilde{C}_1+\tilde{C}_2\tau+\tilde{C}_3\tau^2$ for
some constants $\tilde{C}_1$, $\tilde{C}_2$, $\tilde{C}_3$. \\


\noindent In case 1),
$b(\tau)=C_1+C_2\mathrm{e}^{\lambda\tau}+C_3\mathrm{e}^{-\lambda\tau}+C_4\tau
^2+C_5\tau $ and
$$
[\tilde{T}_1,[\tilde{T_1},\tilde{T}]]-\lambda^2\tilde{T}-\frac{2C_4-\lambda^2C_1}{mc_1}\tilde{T}_1
=-\lambda^2\sum\limits_{j=-\infty}^\infty \Big\{
\rho_j+\frac{C_4\tau_j^2+C_5\tau_j}{mc_1}\Big \}\frac{\partial
}{\partial \tau_j}
$$
is an element in $L_x$. \\

\noindent In case 2),
$b(\tau)=C_1+C_2\tau+C_3\tau^2+C_4\tau^3+C_5\tau^4$ and
$$
\tilde{T}-\frac{C_1}{mc_1}\tilde{T}_1=\sum\limits_{j=-\infty}^\infty
\Big\{\rho_j+\frac{
C_2\tau_j+C_3\tau_j^2+C_4\tau_j^3+C_5\tau_j^4}{mc_1}\Big
\}\frac{\partial}{\partial \tau_j},
$$
belongs to $L_x$.\\
To finish the proof of the Lemma it is enough to show that vector
fields
$$\tilde{T_2}:=
\sum\limits_{j=-\infty}^\infty \{\rho_j+ C_2\tau_j+C_3\tau_j^2+C_4\tau_j^3+
C_5\tau_j^4\}
\frac{\partial}{\partial \tau_j}\, ,
$$
and
$$\tilde{T}_1=\sum\limits_{j=-\infty}^\infty\frac{\partial}{\partial \tau_j}$$
produce an infinite dimensional Lie algebra $L_x$  for any fixed
constants $C_2$, $C_3$, $C_4$ and $C_5$.  One can prove
  it
 by showing  that $L_x$ contains vector fields
 $\sum\limits_{j=-\infty}^\infty
 j^k\frac{\partial}{\partial \tau_j}$, for all $k=1,2,\,\ldots \,\,$.
Note that
$$
[\tilde{T}_1,\tilde{T}_2]=\sum\limits_{j=-\infty}^\infty(-j+C_2+2C_3\tau_j+3C_4\tau_j^2+4C_5\tau_j^3)
\frac{\partial}{\partial \tau_j}.
$$
There are four cases: a) $C_5\ne 0$ and b) $C_5=0, C_4\ne 0$, c)
$C_5=C_4=0$, $C_3\ne 0$ and  d) $C_5=C_4=C_3=0$.\\

\noindent In case a),
$$
[\tilde{T}_1, [\tilde{T_1}, [\tilde{T}_1,
\tilde{T}_2]]]-6C_4\tilde{T}_1=\sum\limits_{j=-\infty}^\infty
24C_5\tau_j\frac{\partial}{\partial \tau_j}=24C_5P_1\in L_x, \quad
P_1=\sum\limits_{j=-\infty}^\infty \tau_j\frac{\partial }{\partial
\tau_j},
$$
$$
[\tilde{T_1}, [\tilde{T}_1,
\tilde{T}_2]]=\sum\limits_{j=-\infty}^\infty
\{2C_3+6C_4\tau_j+12C_5\tau_j^2\}\frac{\partial}{\partial
\tau_j}\in L_x, $$ and  therefore, $$
P_2:=\sum\limits_{j=-\infty}^\infty
\tau_j^2\frac{\partial}{\partial \tau_j}\in L_x,
$$
and
$$
\tilde{T}_3:=[\tilde{T}_1,
\tilde{T}_2]-C_2\tilde{T}_1-2C_3P_1-3C_4P_2=\sum\limits_{j=-\infty}^\infty
(-j+4C_5\tau_j^3)\frac{\partial}{\partial \tau_j}\in L_x.
$$
We have,
$$
J_1:=-\frac{1}{3}([\tilde{T}_3,
P_1]+2\tilde{T}_3)=\sum\limits_{j=-\infty}^\infty
j\frac{\partial}{\partial \tau_j}\in L_x.
$$
Now,
$$
[J_1,[J_1,P_2]]=\frac{1}{2}\sum\limits_{j=-\infty}^\infty
j^2\frac{\partial }{\partial \tau_j} \in L_x.
$$
Assuming $J_k=\sum\limits_{j=-\infty}^\infty
j^k\frac{\partial}{\partial \tau_j}\in L_x$ we have that
$$
J_{k+1}:=\frac{1}{2}[J_1, [J_k,
P_2]]=\sum\limits_{j=-\infty}^\infty
j^{k+1}\frac{\partial}{\partial \tau_j}\in L_x.
$$

\noindent In case b) we have
$$
P_1:=\frac{1}{6C_4}\{[\tilde{T_1}, [\tilde{T}_1,
\tilde{T}_2]]-2C_3\tilde{T_1}\}=\sum\limits_{j=-\infty}^\infty
\tau_j\frac{\partial}{\partial \tau_j}\in L_x
$$
and
$$
\tilde{T}_3=[\tilde{T_1},
\tilde{T_2}]-C_2\tilde{T}_1-2C_3P_1=\sum\limits_{j=-\infty}^\infty
(-j+3C_4\tau_j^2)\frac{\partial}{\partial \tau_j}\in L_x.$$

\noindent We have,
$$
J_1:=-\frac{1}{2}([\tilde{T}_3,
P_1]+\tilde{T}_3)=\sum\limits_{j=-\infty}^\infty
j\frac{\partial}{\partial \tau_j}\in L_x,
$$
and
$$
P_2=\frac{1}{6C_4}(\tilde{T_3}-[\tilde{T}_3,
P_1])=\sum\limits_{j=-\infty}^\infty
\tau_j^2\frac{\partial}{\partial \tau_j}\in L_x.
$$
\noindent As it was shown in the proof of case a),  $J_1$ and
$P_2$ produce an infinite dimensional Lie algebra.\\

\noindent In case c),
$$
\tilde{T}_3=[\tilde{T}_1,
\tilde{T}_2]-C_2\tilde{T}_1=\sum\limits_{j=-\infty}^\infty(-j+2C_3\tau_j)\frac{\partial
}{\partial \tau_j}\in L_x,
$$
$$\tilde{T}_4=[\tilde{T}_3, \tilde{T}_2]=\sum\limits_{j=-\infty}^\infty
(\frac{j(j-1)}{2}-jC_2-2C_3j\tau_j+2C_3^2\tau_j^2 ) \frac{\partial
}{\partial \tau_j}\in L_x.$$

\noindent Also,
$$
\tilde{T}_5=[\tilde{T}_3,\tilde{T}_4]
=2C_3\sum\limits_{j=-\infty}^\infty\Big(\frac{j(j+1)}{2}+C_2j-2C_3j\tau_j+2C_3^2\tau_j^2\Big)\frac{\partial
}{\partial \tau_j}\in L_x.
$$
Since $\tilde{T}_4$ and $\tilde{T}_5$ both belong to $L_x$ then
either
$$
c) (i)\quad J_1=\sum\limits_{j=-\infty}^\infty j\frac{\partial
}{\partial \tau_j}\in L_x, \qquad
\tilde{T}_6=\sum\limits_{j=-\infty}^\infty(\frac{j^2}{2}-2C_3j\tau_j+2C_3^2\tau_j^2)\frac{\partial
}{\partial \tau_j}\in L_x,
$$
or
$$
c) (ii) \quad C_2=-\frac{1}{2}, \quad
\tilde{T}_6=\sum\limits_{j=-\infty}^\infty(\frac{j^2}{2}-2C_3j\tau_j+2C_3^2\tau_j^2)\frac{\partial
}{\partial \tau_j}\in L_x.
$$
\noindent In case c) (i),
$$
P_1=\frac{1}{4C_3^2}\{[\tilde{T}_1,
\tilde{T}_6]+2C_3J_1\}=\sum\limits_{j=-\infty}^\infty
\tau_j\frac{\partial}{\partial \tau_j}\in L_x.
$$
\noindent Since
$$
[P_1, \tilde{T}_6]=\sum_{j=-\infty}^\infty
(-\frac{j^2}{2}+2C_3^2\tau_j^2)\frac{\partial}{\partial \tau_j},
$$
\noindent and
$$
[P_1,[P_1, \tilde{T}_6]]=\sum_{j=-\infty}^\infty
(\frac{j^2}{2}+2C_3^2\tau_j^2)\frac{\partial}{\partial \tau_j}
$$
both belong to $L_x$ then
$$
J_2=\sum_{j=-\infty}^\infty j^2\frac{\partial}{\partial \tau_j}\in
L_x, \qquad P_2=\sum_{j=-\infty}^\infty
\tau_j^2\frac{\partial}{\partial \tau_j}\in L_x,
$$
$P_2$ and $J_1$ generate an infinite dimensional Lie algebra.\\

\noindent In case c) (ii),
$$
\tilde{T}_1=\sum_{j=-\infty}^\infty\frac{\partial }{\partial
\tau_j} , \quad
\tilde{T}_2=\sum_{j=-\infty}^\infty\Big(C_3\tau_j^2-\frac{1}{2}\tau_j+\rho_j\Big)\frac{\partial
}{\partial \tau_j}.
$$
Note that the Lie algebra generated by the vector fields
$$
\tilde{T_2^*}=\tilde{T}_2-\Big(C_3\tau^2-\frac{1}{2}\tau\Big)\tilde{T}_1
=d(\tau, \tau_1)\frac{\partial}{\partial \tau_1}-d(\tau_{-1},
\tau)\frac{\partial}{\partial \tau_{-1}}+(d(\tau, \tau_1)+d(\tau_1,
\tau_{2}))\frac{\partial}{\partial \tau_{2}}+\ldots$$ and
$$\tilde{T}_1=\sum_{j=-\infty}^\infty\frac{\partial }{\partial
\tau_j}
$$
is infinite dimensional. It can be proved by comparing this
algebra with the infinite dimensional characteristic Lie algebra
of the chain
\begin{equation}\label{fine}
t_{1x}=t_x +C_3(t_1^2-t^2)-\frac{1}{2}(t_1+t).
\end{equation}
Indeed, the Lie algebra $L_{x1}$ for (\ref{fine}) is generated by
the operators (\ref{tildeX}) and (\ref{tildeY}) with
$d(t,t_1)=C_3(t_1^2-t^2)-\frac{1}{2}(t_1+t)$. To keep standard
notations we put $a(\tau)=-2C_3\tau-1$ and
$b(\tau)=C_3\tau^2+\frac{1}{2}\tau.$ Note that since $C_3\neq0$
function $a(\tau)$ is not a constant. It follows from Theorem 3
proved below that the characteristic Lie algebras $L_x$ (and
therefore algebra $L_{x1}$) for equation (\ref{fine}) is of
infinite dimension. Thus, in case c) (ii) we also have an infinite
dimensional Lie algebra $L_x$.

\noindent In case d),
$$
\tilde{T}_2=\sum\limits_{j=-\infty}^\infty
(-\tau-\tau_1-\ldots-\tau_{j-1}+C_2\tau_j)\frac{\partial
}{\partial \tau_j}\in L_x.
$$
Then
$$J_1=c_2\tilde{T_1}-[\tilde{T}_1, \tilde{T}_2]=\sum\limits_{j=-\infty}^\infty
j\frac{\partial }{\partial \tau_j}\in L_x,
$$
and
$$
J_2=-2\Big([J_1,
\tilde{T}_2]-\Big(\frac{1}{2}+C_2\Big)J_1\Big)=\sum\limits_{j=-\infty}^\infty
j^2\frac{\partial }{\partial \tau_j}\in L_x.
$$
Assuming that $J_k$, $1\leq k\leq n$ belong to $L_x$, by
considering $[J_n,\tilde{T}_2]$ one may show that
$J_{n+1}=\sum\limits_{j=-\infty}^\infty j^{k+1}\frac{\partial
}{\partial \tau_j}\in L_x$. It implies $L_x$ is of infinite
dimension. $\Box$\\


\noindent Let us introduce new variables
$$
w_j=\ln\Big(\tau_j+\frac{c_1}{c_0}\Big).
$$
Vector fields $T_1$ and $T$ in variables $w_j$ can be rewritten as
$$
T_1=c_0\sum\limits_{j=-\infty}^\infty \frac{\partial}{\partial
w_j}=c_0S_0\, ,
$$
$$T=-tc_0S_0+c_0\sum\limits_{j=-\infty}^\infty \{\tilde{\rho}_j+\tilde{b}(w_j)\}\frac{\partial}{\partial
w_j}=-c_0tS_0+c_0S_1,
$$
where
$$
S_0=\sum\limits_{j=-\infty}^\infty \frac{\partial}{\partial w_j},
\quad S_1=\sum\limits_{j=-\infty}^\infty
\{\tilde{\rho}_j+\tilde{b}(w_j)\}\frac{\partial}{\partial w_j},
$$
$$
\tilde{\rho_j}=\left\{\begin{array}{cl}\sum\limits_{k=0}^{j-1}(\mathrm{e}^{w_k}-\tilde{c}),
& {\rm{if}}
\quad j\geq 1;\\
0, & {\rm{if}}
\quad j=0;\\
-\sum\limits_{k=j}^{-1}(\mathrm{e}^{w_k}-\tilde{c}), & {\rm{if}}
\quad j\leq -1,
\end{array}
\right. \quad \tilde{c}=\frac{c_1}{c_0}, \quad
\tilde{b}(w_j)=-\frac{1}{m}\Big(\frac{b(\tau_j)}{c_0\tau_j+c_1}\Big)\,
.
$$
We have
$$
DS_0D^{-1}=S_0,\quad DS_1D^{-1}=S_1-(\mathrm{e}^w-\tilde{c})S_0.
$$

\noindent These lemmas allow one to prove the following Theorem.

\begin{theorem}
If equation
$$
t_{1x}=t_x+a(\tau)t^m+b(\tau)t^{m-1}+\ldots, \quad m\geq 1
$$
admits a nontrivial $x$-integral, then\\
(1) $a(\tau)=c_0\tau$, $b(\tau)=c_2\tau^2+c_3\tau$, where $c_0$, $c_2$, $c_3$ are some constants.\\
(2) $m=1$.
\end{theorem}

\noindent{\bf Proof}. Consider the case (1).  Define vector field
$$Q=[S_0,[S_0,S_1]]-[S_0,S_1]=\sum\limits_{j=-\infty}^\infty (\tilde{b''}(w_j)-
\tilde{b'}(w_j))\frac{\partial}{\partial w_j}.$$ By Lemma
\ref{LemmaAux2}, $\tilde{b''}(w)-\tilde{b'}(w)=C$ for some
constant $C$. Thus, $\tilde{b}(w)=C_0+C_1\mathrm{e}^w+C_2w$ for
some constants $C_1$, $C_2$, $C_0$. Consider vector fields
$$P=(C_2-C_0)S_0+S_1-[S_0,S_1]=\sum\limits_{j=-\infty}^\infty
(C_2w_j+\tilde{c}j)\frac{\partial}{\partial w_j},$$
$$
R=[S_0,[S_0,S_1]]=\sum\limits_{j=1}^\infty\left\{
\Big(\sum_{k=1}^j \mathrm{e}^{w_k}\Big)+C_1
\mathrm{e}^{w_j}\right\}\frac{\partial}{\partial
w_j}+C_1\mathrm{e}^w\frac{\partial}{\partial
w}-\sum\limits_{j=-\infty}^{-1} \left\{\Big(\sum_{k=j}^{-1}
\mathrm{e}^{w_k}\Big)+C_1
\mathrm{e}^{w_j}\right\}\frac{\partial}{\partial w_j},$$
$$
R_1=[P,R], \quad R_{n+1}=[P, R_n], \quad n\geq 1.
$$
Then
\begin{eqnarray*}
R_{n}&=&\sum_{j\geq 0}\{ \mathrm{e}^{w_j}(C_1C_2^n
w_j^n+P_{n,j})+r_{n,j}(w,w_1, \ldots, w_{j-1})\}
\frac{\partial}{\partial w_j}\\ &+& \sum_{j\leq -1}\{
\mathrm{e}^{w_j}((C_1-1)C_2^n
w_j^n+P_{n,j})+r_{n,j}(w_{-1},w_{-2}, \ldots, w_{j+1})\}
\frac{\partial}{\partial w_j},
\end{eqnarray*}
where $P_{n,j}=P_{n,j}(w_j, j)$ is a polynomial of degree $n-1$
whose coefficients depend on $j$, $r_{n,j}$ are the functions that
do not depend on $w_j$. Since all vector fields $R_n$ belong to a
finite dimensional Lie algebra $L_x$ then $C_1C_2=(C_1-1)C_2=0$,
or the same $C_2=0$. Therefore,
$$
\tilde{b}(w)=C_0+C_1\mathrm{e}^w\, .$$ Since $C_2=0$, then
$$P=\tilde{c}\sum\limits_{j=-\infty}^\infty
j\frac{\partial}{\partial w_j},$$
$$
R=\sum\limits_{j=1}^\infty \left\{\Big(\sum_{k=1}^j
\mathrm{e}^{w_k}\Big)+C_1
\mathrm{e}^{w_j}\right\}\frac{\partial}{\partial
w_j}+C_1\mathrm{e}^w\frac{\partial}{\partial
w}-\sum\limits_{j=-\infty}^{-1} \left\{\Big(\sum_{k=j}^{-1}
\mathrm{e}^{w_k}\Big)+C_1
\mathrm{e}^{w_j}\right\}\frac{\partial}{\partial w_j}$$ and
\begin{eqnarray*}
R_n&=&\tilde{c}^n\sum\limits_{j=1}^\infty
\{\mathrm{e}^{w_1}+2^{n}\mathrm{e}^{w_2}+(j-1)^n\mathrm{e}^{w_{j-1}}+j^nC_1
\mathrm{e}^{w_j}\}\frac{\partial}{\partial w_j}\\
&-&\tilde{c}^n\sum\limits_{j=-\infty}^{-1}
\{(-1)^n\mathrm{e}^{w_{-1}}+(-2)^{n}\mathrm{e}^{w_{-2}}+(j)^n\mathrm{e}^{w_{j}}+j^nC_1
\mathrm{e}^{w_j}\}\frac{\partial}{\partial w_j}.
\end{eqnarray*}
\noindent Again, vector fields $R_n$ belong to a finite
dimensional Lie algebra only if $\tilde{c}=0$, or the same
$c_1=0$. It implies that
$$
 a(\tau)=c_0\tau, \quad b(\tau)= c_2\tau^2+c_3\tau.
$$
Consider the case (2). Assume contrary, that is $m\geq 2$. Then the
following vector field
\begin{eqnarray*}
&&\frac{1}{m!}ad_{\tilde{X}}^{m-2}(\tilde{Y})
=Y_{\frac{1}{2}a(\tau)t^2+\frac{1}{m}b(\tau)t+\frac{1}{m(m-1)}c(\tau)}\\
&&=
 -\sum\limits_{j=-\infty}^\infty
(\frac{1}{2}a(\tau_j)t_j^2+\frac{1}{m}b(\tau_j)t_j+\frac{1}{m(m-1)}c(\tau_j))
\frac{\partial}{\partial \tau_j}\\
&&=-\sum\limits_{j=-\infty}^\infty
(\frac{1}{2}a(\tau_j)(t+\rho_j)^2+\frac{1}{m}b(\tau_j)(t+\rho_j)+\frac{1}{m(m-1)}c(\tau_j))
\frac{\partial}{\partial \tau_j}
-\frac{t^2}{2}\sum\limits_{j=-\infty}^\infty
a(\tau_j)\frac{\partial}{\partial \tau_j}\\
&&-t\sum\limits_{j=-\infty}^\infty
\{a(\tau_j)\rho_j+\frac{1}{m}b(\tau_j)\}\frac{\partial}{\partial
\tau_j}- \sum\limits_{j=-\infty}^\infty
\{\frac{1}{2}a(\tau_j)\rho^2_j+\frac{1}{m}b(\tau_j)+\frac{1}{m(m-1)}c(\tau_j)\}
\frac{\partial}{\partial \tau_j}
\end{eqnarray*}
is in $L_x$.  In variables $w_j=\ln\tau_j$,
$$
\frac{1}{m!}ad_{\tilde{X}}^{m-2}(\tilde{Y})=-\frac{t^2}{2}c_0S_0+tc_0S_1-c_0S_2,$$
where
$$
S_2=\sum\limits_{j=-\infty}^\infty
\{\frac{1}{2}\tilde{\rho}^2_j-\tilde{b}(w_j)\tilde{\rho}_j+\tilde{c}(w_j)\}
\frac{\partial}{\partial w_j}, \quad
\tilde{c}(w_j)=\frac{c(\tau_j)}{m(m-1)\tau_j}.
$$
The vector fields $S_0$ and $S_1$ are as in Lemma \ref{LemmaAux2}.
We have,
$$
[S_0, S_2]=2S_2+C_0S_1+P, \quad P= \sum\limits_{j=-\infty}^\infty
r(w_j)\frac{\partial}{\partial w_j}, \quad
r(w)=\tilde{c}'(w)-2\tilde{c}(w)-C_0\tilde{b}(w).
$$
Construct the sequence
$$S_3=[S_1,S_2], \quad S_{n+1}=[S_1, S_n], \quad n\geq 2.
$$
One can prove by induction on $n$ that
$$
[S_0, S_n]=nS_n+\sum_{k=0}^{n-1}\nu_{n,k}S_k,
$$
and
$$
DS_nD^{-1}=S_n+\left\{ \frac{n(n-1)}{2}-1
\right\}\mathrm{e}^wS_{n-1}+\sum_{k=0}^{n-2}\eta(n,k)S_k,\quad
n\geq 3.
$$
Since $L_x$ is of finite dimension then there exists a natural
number $N$ such that
$$
S_{N+1}=\mu_NS_N+\mu_{N-1}S_{N-1}+\ldots+\mu_0S_0.
$$
Then
$$
DS_{N+1}D^{-1}=D(\mu_N)DS_ND^{-1}+D(\mu_{N-1})DS_{N-1}D^{-1}+\ldots+D(\mu_0)DS_0D^{-1}.
$$
On the other hand,
$$DS_{N+1}D^{-1}=S_{N+1}+\left\{\frac{(N+1)N}{2}-1\right\}\mathrm{e}^wS_N+\ldots\,.
$$
We compare the coefficients before $S_N$ and have two equations.
$$
D(\mu_N)=\mu_N+\left\{\frac{(N+1)N}{2}-1\right\}\mathrm{e}^w,
\quad N\geq 2,
$$
and
$$
D(\mu_1)=\mu_1+\mathrm{e}^w,\quad  N=1.
$$
Both equation are contradictory. Therefore, our assumption that
$m\geq 2$ was wrong. $\Box$

\section{Nonzero root}

\begin{lemma}
\label{Lemma22} Assume equation (\ref{main}) admits a nontrivial
$x$-integral. Then the characteristic polynomial of the equation
(\ref{DifEqalpha1}) can have only simple nonzero roots.
\end{lemma}

\noindent{\bf{Proof}}.
 Assume that $m_1\geq 2$.
 Introduce polynomials
 $$
 \Lambda^{(2)}_{\alpha_1}(\lambda)=\frac{\Lambda(\lambda)}{(\lambda-\alpha_1)^2},\quad
 \Lambda_{\alpha_1}(\lambda)=\frac{\Lambda(\lambda)}{(\lambda-\alpha_1)}.
$$
 Consider vector fields
 $$
 S_0^*=\Lambda^{(2)}_{\alpha_1}(ad_{\tilde{X}})Y_d=Y_{A(\tau)\mathrm{e}^{\alpha_1 t}}\,
 $$
 $$
 S_1^*=\Lambda_{\alpha_1}(ad_{\tilde{X}})Y_d=Y_{(A(\tau)t+B(\tau))\mathrm{e}^{\alpha_1 t}}
 $$
  from the the Lie algebra $L_x$.

\noindent In variables $\tau_j=t_j-t_{j+1}$, vector fields $S_0^*$
and $S_1^*$ become
$$
S^*_0=-\mathrm{e}^{\alpha_1t}\sum\limits_{j=-\infty}^\infty
A(\tau_j)\mathrm{e}^{\alpha_1 \rho_j}\frac{\partial}{\partial
\tau_j}=-\mathrm{e}^{\alpha_1t}S_0,
$$
$$
S_1^*=-t\mathrm{e}^{\alpha_1t}S_0-\mathrm{e}^{\alpha_1t}\sum\limits_{j=-\infty}^\infty
\{A(\tau_j)\rho_j+B(\tau_j)\}\mathrm{e}^{\alpha_1
\rho_j}\frac{\partial}{\partial
\tau_j}=-t\mathrm{e}^{\alpha_1t}S_0-\mathrm{e}^{\alpha_1t}S_1,
$$
with $S_0=\sum\limits_{j=-\infty}^\infty
A(\tau_j)\mathrm{e}^{\alpha_1 \rho_j}\frac{\partial}{\partial
\tau_j}$ and  $S_1=\sum\limits_{j=-\infty}^\infty
\{A(\tau_j)\rho_j+B(\tau_j)\}\mathrm{e}^{\alpha_1
\rho_j}\frac{\partial}{\partial \tau_j}.$

\noindent Direct calculations show that
$$
DS_0D^{-1}=\mathrm{e}^{\alpha_1\tau}S_0, \quad
DS_1D^{-1}=\mathrm{e}^{\alpha_1\tau }S_1+\tau
\mathrm{e}^{\alpha_1\tau }S_0.
$$
Define the sequence
$$
S_2=[S_0, S_1], \quad S_{n+1}=[S_0, S_n], \quad n\geq 2.
$$
One can easily show that
$$
DS_2D^{-1}=\mathrm{e}^{2\alpha_1\tau}S_2+\alpha_1\mathrm{e}^{2\alpha_1\tau}A(\tau)S_1+\mathrm{e}^{2\alpha_1\tau}(A(\tau)-\alpha_1B(\tau))S_0\,
.
$$
It can be proved by induction on $n$ that
$$
DS_nD^{-1}=\mathrm{e}^{n\alpha_1\tau}S_n+\alpha_1\frac{n(n-1)}{2}\mathrm{e}^{n\alpha_1\tau}A(\tau)S_{n-1}+\sum\limits_{k=0}^{n-2}
\gamma(n,k)S_k\, .
$$
Since the dimension of $L_x$ is finite and $S_0$, $S_1$, $\ldots$
are elements of $L_x$ then there exists a natural number $N$ such
that
$$
S_{N+1}=\mu_NS_N+\mu_{N-1}S_{N-1}+\ldots +\mu_0S_0,
$$
and $S_0$, $S_1$, $\ldots$, $S_N$ are linearly independent.
Therefore,
$$
DS_{N+1}D^{-1}=D(\mu_N)DS_ND^{-1}+D(\mu_{N-1})DS_{N-1}D^{-1}+\ldots +
D(\mu_0)DS_0D^{-1}.
$$
On the other hand,
$$
DS_{N+1}D^{-1}=\mathrm{e}^{(N+1)\alpha_1\tau}S_{N+1}+\alpha_1\frac{(N+1)N}{2}\mathrm{e}^{(N+1)\alpha_1\tau}A(\tau)S_N+\sum\limits_{k=0}^{N-1}
\gamma(N+1,k)S_k.
$$
By comparing the coefficients before $S_N$ in the last two
equations we have
$$
\mathrm{e}^{(N+1)\alpha_1\tau}\mu_N+\frac{\alpha_1
(N+1)N}{2}\mathrm{e}^{(N+1)\alpha_1\tau}
A(\tau)=D(\mu_N)\mathrm{e}^{N\alpha_1\tau}.
$$
It follows at once that $\mu_N$ is a constant and then
$$
A(\tau)=C(\mathrm{e}^{-\alpha_1\tau}-1), \quad
C=\frac{2\mu_N}{\alpha_1N(N+1)}.
$$
Let us construct a new infinite sequence of vector fields
belonging to $L_x$, enumerated by a multi-index.
$$
T_0:=S_1,\quad T_1:=S_0,\quad T_2=[S_1, T_1], \quad
T_{n+1}=[S_1,T_n], \quad n\geq 2, \quad T_{n,0}=[S_0, T_n] ,
$$
$$
T_{n, 0, i_1, \ldots, i_{n-1}, i_n}=[S_{i_n},T_{n, 0, i_1, \ldots,
i_{n-1}}], \quad i_j\in\{0;1\}.
$$
Direct calculations show that
$$
DT_2D^{-1}=\mathrm{e}^{2\alpha_1\tau}T_2+\mathrm{e}^{2\alpha_1\tau}(\alpha_1B-A)T_1-\alpha_1\mathrm{e}^{2\alpha_1\tau}AT_0,
$$
$$
DT_3D^{-1}=\mathrm{e}^{3\alpha_1\tau}T_3+\mathrm{e}^{3\alpha_1\tau}(3\alpha_1B-A+3\alpha_1\tau
A)T_2+\tau \mathrm{e}^{3\alpha_1\tau }T_{2,0}
+\sum\limits_{m(\beta)<2 }\nu(3, \beta)T_\beta\, .
$$
Here and below we use functions $m=m(\beta)$ and $l=l(\beta)$
defined in Section 3. It can be proved by induction on $n$ that
$$
DT_nD^{-1}=\mathrm{e}^{n\alpha_1\tau}T_n+\mathrm{e}^{n\alpha_1\tau}\{c_nB-A+c_n\tau
A \}T_{n-1}+\tau \mathrm{e}^{n\alpha_1\tau}
\sum\limits_{m(\beta)=n-1, l(\beta)=1}\nu^*(n, \beta)T_\beta+
\sum\limits_{m(\beta)\leq n-2}\nu(n, \beta)T_\beta\, ,
$$
\noindent where
$$
c_n=\frac{\alpha_1 n(n-1)}{2},
$$
and $\nu^*(n, \beta)$ are constants for any $\beta$ with
$m(\beta)=n-1$ and $l(\beta)=1$.

\noindent In general, for any $\gamma$,
$$
DT_\gamma
D^{-1}=\mathrm{e}^{(m(\gamma)+l(\gamma))\alpha_1\tau}T_\gamma
+\sum\limits_{m(\beta)\leq m(\gamma)-1}\nu(\gamma, \beta)T_\beta\,
.
$$
Among the vector fields $T_\beta$ we choose a system $P$ of linearly
independent vector fields in such a way that for some natural number
$N$
\\
(i) $T_k\in P, \quad k\leq N$,\\
(ii) $m(\beta)\leq N$ for any $T_\beta \in P$. \\
(iii) for any $T_{\gamma}$ with $m(\gamma)\leq N$ we have
$T_\gamma=\sum\limits_{T_\beta\in P, m(\beta)\leq
m(\gamma)}\mu(\gamma, \beta)T_\beta$. Also

 $T_{N+1}=\mu(N+1, N)T_N+\sum\limits_{T_\beta\in P}\mu(N+1, \beta)T_\beta.$\\
(iv) for any $T_\gamma\notin P$ with $m(\gamma)=N$ and $l(\gamma)=1$, we have $\mu(\gamma, N)=0$.\\
Indeed,
$$
DT_\gamma
D^{-1}=D(\mu(\gamma,N))DT_ND^{-1}+\sum\limits_{T_\beta\in P,
\beta\ne N}D(\mu(\gamma, \beta))DT_\beta D^{-1}\, .
$$
On the other hand,
\begin{eqnarray*}
DT_\gamma
D^{-1}&=&\mathrm{e}^{(m(\gamma)+l(\gamma))\alpha_1\tau}T_\gamma
+\sum\limits_{m(\beta)\leq N-1}\nu(\gamma, \beta)T_\beta\\
&=&\mathrm{e}^{(N+1)\alpha_1\tau} \{\mu(\gamma,
N)T_N+\sum_{T_\beta\in P, m(\beta)\leq N, \beta\ne N}\mu(\gamma,
\beta)T_\beta \} +\sum\limits_{m(\beta)\leq N-1}\nu(\gamma,
\beta)T_\beta.
\end{eqnarray*}
By comparing the coefficients before $T_N$ we have
 $$
 \mathrm{e}^{(N+1)\alpha_1\tau}\mu(\gamma, N)=D(\mu(\gamma, N))\mathrm{e}^{N\alpha_1\tau}
 $$
 that proves $\mu(\gamma, N)=0$ for any $\gamma$ with $m(\gamma)=N$ and $l(\gamma)=1$.
 \noindent
 We have,
$$
T_{N+1}=\mu_NT_N+\sum\limits_{T_\beta\in P}\mu(N+1, \beta)T_\beta,
$$
here $\mu_N=\mu(N+1, N)$. Then
$$
DT_{N+1}D^{-1}=D(\mu_N)DT_ND^{-1}+\sum\limits_{T_\beta\in
P}D(\mu(N+1, \beta))DT_\beta D^{-1}.
$$
We continue and have,
\begin{eqnarray*}
&&\mathrm{e}^{(N+1)\alpha_1\tau}\{
\mu_NT_N+\sum\limits_{T_\beta\in P}\mu(N+1, \beta)T_\beta\}+
\mathrm{e}^{(N+1)\alpha_1\tau}\{c_{N+1}B-A+c_{N+1}\tau A\}T_N\\
&&+\tau \mathrm{e}^{(N+1)\alpha_1\tau}\sum\limits_{m(\beta)=N,
l(\beta)=1} \nu^*(N+1, \beta)T_\beta+\sum\limits_{m(\beta)\leq
N-1} \nu(N+1,
\beta)T_\beta\\
&&=D(\mu_N)\{\mathrm{e}^{N\alpha_1\tau}T_N+\sum\limits_{m(\beta)\leq
N-1}\nu(N, \beta)T_\beta\}\\
&&+\sum\limits_{T_\beta\in P}D(\mu(N+1,
\beta))\{\mathrm{e}^{(m(\beta)+l(\beta))\alpha_1 \tau}T_\beta+
\sum\limits_{m(r)\leq N-1}\nu(\beta, r)T_r\}.
\end{eqnarray*}
We compare the coefficients before $T_N$ and get
$$
\mathrm{e}^{(N+1)\alpha_1\tau}\mu_N+\mathrm{e}^{(N+1)\alpha_1\tau}\{c_{N+1}B-A+c_{N+1}\tau
A\}=\mathrm{e}^{N\alpha_1\tau}D(\mu_N).$$

\noindent Note that, by property (iv), we do not have term $\tau
\mathrm{e}^{(N+1)\alpha_1\tau}$ in the left side of the last
equality. Thus, using the expression for
$A(\tau)=C(\mathrm{e}^{-\alpha_1\tau}-1)$ and the fact that
$\mu_N$ is a constant, we have
$$
B(\tau)=C_1A+C_2\tau
A=C_1(\mathrm{e}^{-\alpha_1\tau}-1)+C_2\tau(\mathrm{e}^{-\alpha_1\tau
}-1)\, ,
$$
\noindent where
$$
C_1=\frac{\mu_N}{Cc_{N+1}}+\frac{1}{c_{N+1}}, \quad C_2=-1 .
$$
We introduce new vector fields
$$
\tilde{S_0}=\frac{1}{C}S_0=(\mathrm{e}^{-\alpha_1
\tau}-1)\frac{\partial}{\partial \tau}+\ldots, \quad
\tilde{S_1}=\frac{1}{C}S_1+\frac{C_1}{C}S_0=\tau(\mathrm{e}^{-\alpha_1\tau}-1)\frac{\partial}{\partial
\tau}+\ldots\,.
$$
$$
\tilde{S_2}=[\tilde{S}_0, \tilde{S}_1], \quad
\tilde{S}_{n+1}=[\tilde{S}_0, \tilde{S}_n], \quad n\geq 2.
$$
We have,
$$
D\tilde{S}_0D^{-1}=\mathrm{e}^{\alpha_1\tau}\tilde{S}_0, \quad
D\tilde{S}_1D^{-1}=\mathrm{e}^{\alpha_1\tau}\tilde{S}_1-\tau
\mathrm{e}^{\alpha_1\tau}\tilde{S}_0,
$$
$$
D\tilde{S}_nD^{-1}=\sum\limits_{k=0}^n\tilde{\gamma}(n,
k)\tilde{S}_k, \quad
\tilde{\gamma}(n,n)=\mathrm{e}^{n\alpha_1\tau}\, ,
$$
where $\tilde{\gamma}(n,k)$ are functions of $\tau$ only. Since
all vector fields $\tilde{S}_k$ belong to a finite dimensional Lie
algebra $L_x$, then there exists such a natural number $M$ that
\begin{equation}\label{tilde}
\tilde{S}_{M+1}=\tilde{\mu}_M\tilde{S}_M+\ldots
+\tilde{\mu}_0\tilde{S}_0,
\end{equation}
and $\tilde{S}_0$, $\ldots$, $\tilde{S}_M$ are linearly
independent. Then
$$
D\tilde{S}_{M+1}D^{-1}=D(\tilde{\mu}_M)D\tilde{S}_MD^{-1}+\ldots
+D(\tilde{\mu}_0)D\tilde{S}_0D^{-1},
$$
and
$$
\tilde{\gamma}(M+1, M+1)\{
\tilde{\mu}_M\tilde{S}_M+\ldots+\tilde{\mu}_0
\tilde{S}_0\}+\sum\limits_{k=0}^M\tilde{\gamma}(M+1,
k)\tilde{S}_k= D(\tilde{\mu}_N)\{ \tilde{\gamma}(M,
M)\tilde{S}_M+\ldots\}+\ldots\,.
$$
 By comparing the coefficients before $\tilde{S}_M$,  we have
 $$
 \mathrm{e}^{(M+1)\alpha_1\tau}\tilde{\mu}_M+\tilde{\gamma}(M+1,M)=D(\tilde{\mu}_M)\mathrm{e}^{M\alpha_1\tau}
 $$
 that implies that $\tilde{\mu}_M$ is a constant. In the same way, by comparing the coefficients before
 $\tilde{S}_{M-1}$, and then before $\tilde{S}_{M-2}$, and so on, one can show that all coefficients
 $\tilde{\mu}_k$ are constants.

\noindent One can show by induction on $n$ that for $n\geq 2$,
 $$
 \tilde{S}_n=\{\alpha_1^{n-2}(-1)^{n-2}(n-2)!\mathrm{e}^{-n\alpha_1\tau}+\sum\limits_{k=0}^{n-1}
 r(n,k)\mathrm{e}^{-\alpha_1k\tau}\}\frac{\partial}{\partial \tau}+\ldots\, ,
 $$
 where
 $r(n,k)$ are some constants. Return to equality (\ref{tilde}) with constant coefficients $\tilde{\mu}_k$ and compare the coefficients before $\frac{\partial}{\partial \tau}$:
 \begin{eqnarray*}
 &&\alpha_1^{M-1}(-1)^{M-1}(M-1)!\mathrm{e}^{-(M+1)\alpha_1\tau}+\sum\limits_{k=0}^M r(M+1, k)\mathrm{e}^{-\alpha_1
 k\tau}\\
 &&=\tilde{\mu}_M
 \Big(\alpha_1^{M-2}(-1)^{M-2}(M-2)!\mathrm{e}^{-M\alpha_1\tau}+\sum\limits_{k=0}^{M-1} r(M,
 k)\mathrm{e}^{-\alpha_1k\tau}\Big)+\ldots+\tilde{\mu}_0(\mathrm{e}^{-\alpha_1\tau}-1).
 \end{eqnarray*}
 The last equality fails to be true. It shows that our assumption that multiplicity $m_1$ of a nonzero root $\alpha_1$ can be 2 or more was wrong.
 $\Box$

If the characteristic polynomial of (\ref{DifEqalpha1}) has only
one nonzero root $\alpha$, then $d(t,t_1)=A(t-t_1)e^{\alpha t}$.
In this case equation (\ref{main}) admits a nontrivial
$x$-integral (see Introduction, Theorem 3). In the next section we
consider a case when the characteristic polynomial of
(\ref{DifEqalpha1}) has at least two nonzero roots.

 \section{Two nonzero roots}

 Let $\alpha$ and $\beta$ be two nonzero roots. Consider the vector fields
 $$
 S_0=\sum\limits_{j=-\infty}^\infty A(\tau_j)\mathrm{e}^{\alpha \rho_j}\frac{\partial }{\partial \tau_j}, \quad
 S_1=\sum\limits_{j=-\infty}^\infty B(\tau_j)\mathrm{e}^{\beta \rho_j}\frac{\partial }{\partial \tau_j}
 $$
 from the Lie algebra $L_x$,
 and construct a new sequence of vector fields
 $$
 S_2=[S_0, S_1], \quad S_{n+1}=[S_0, S_n],\quad n\geq 1.
 $$
 We have,
 $$
 DS_0D^{-1}=\mathrm{e}^{\alpha\tau}S_0, \quad DS_1D^{-1}=\mathrm{e}^{\beta\tau}S_1\, ,
 $$
  $$DS_2D^{-1}=\mathrm{e}^{(\alpha+\beta)\tau}S_2+\beta A\mathrm{e}^{(\alpha+\beta)\tau}S_1-\alpha B\mathrm{e}^{(\alpha+\beta)\tau}S_0\, .
  $$
 In general, for any $n\geq 3$,
 $$
 DS_nD^{-1}=\mathrm{e}^{((n-1)\alpha+\beta)\tau}\{S_n+(c_n\alpha+d_n\beta)AS_{n-1}+(p_nA'+q_nA)AS_{n-2}+
 \sum\limits_{k=0}^{n-2}\nu(n, k)S_k\}\, ,
 $$
 where
 $$
 c_n=\frac{(n-1)(n-2)}{2}\, ,\quad d_n=n-1, \quad p_{n+1}=\frac{n(n-1)}{2}\left\{ \frac{n-2}{3}\alpha+\beta\right\}\, , \quad n\geq 2 ,
 $$
 $$
 q_{n+1}=\frac{n(n-2)(n-1)(3n-1)}{24}\alpha^2+\frac{(n-1)^2n}{2}\alpha\beta+\frac{n(n-1)}{2}\beta^2\, ,\quad n\geq
 2.
 $$
 Let us consider  a particular case when
 \begin{equation}\label{S2asli}
 S_2=\mu_0S_0+\mu_1S_1.
\end{equation}
 We have,
 \begin{eqnarray*}
 DS_2D^{-1}&=&D(\mu_0)\mathrm{e}^{\alpha\tau}S_0+D(\mu_1)\mathrm{e}^{\beta\tau}S_1=
 \mathrm{e}^{(\alpha+\beta)\tau}S_2+\beta A\mathrm{e}^{(\alpha+\beta)\tau}S_1-\alpha
 B\mathrm{e}^{(\alpha+\beta)\tau}S_0\\
 &=&
 \mathrm{e}^{(\alpha+\beta)\tau}\{\mu_0S_0+\mu_1S_1\}+\beta A\mathrm{e}^{(\alpha+\beta)\tau}S_1-\alpha
 B\mathrm{e}^{(\alpha+\beta)\tau}S_0.
 \end{eqnarray*}
 Comparing coefficients before $S_0$ and $S_1$ produces the following two equations
 $$
 \mathrm{e}^{(\alpha+\beta)\tau}\mu_0-\alpha B\mathrm{e}^{(\alpha+\beta)\tau}=D(\mu_0)\mathrm{e}^{\alpha\tau}, \quad
 \mathrm{e}^{(\alpha+\beta)\tau}\mu_1+\beta
 A\mathrm{e}^{(\alpha+\beta)\tau}=D(\mu_1)\mathrm{e}^{\beta\tau}.
 $$
 It follows that $\mu_0$, $\mu_1$ are constants and
 $$
 B(\tau)=-\frac{\mu_0}{\alpha}(\mathrm{e}^{-\beta\tau}-1), \quad
 A(\tau)=\frac{\mu_1}{\beta}(\mathrm{e}^{-\alpha\tau}-1).
 $$
And finally, comparing coefficients before
$\frac{\partial}{\partial \tau}$ in equation (\ref{S2asli})
implies that $\alpha=-\beta$.

Let us return to the general case. Since $L_x$ is of finite
dimension then there exists such number $N$ that $S_0$, $S_1$,
$\ldots$, $S_N$ are linearly independent and
$$
S_{N+1}=\mu_NS_N+\mu_{N-1}S_{N-1}+\ldots+\mu_0S_0.
$$
Then
$$
DS_{N+1}D^{-1}=D(\mu_N)DS_ND^{-1}+D(\mu_{N-1})DS_{N-1}D^{-1}+\ldots+D(\mu_0)DS_0D^{-1}
$$
and therefore,
$$
\mathrm{e}^{(N\alpha
+\beta)\tau}\{(\mu_NS_N+\mu_{N-1}S_{N-1}+\ldots)+A(c_{N+1}\alpha+d_{N+1}\beta)S_N+A(p_{N+1}A'+q_{N+1}A)S_{N+1}+\ldots\}=
$$
$$
D(\mu_N)\{ \mathrm{e}^{((N-1)\alpha+\beta)\tau}(S_N+A(c_N\alpha
+d_N\beta)S_{N-1}+\ldots)\}+D(\mu_{N-1})\{
\mathrm{e}^{((N-2)\alpha+\beta)\tau}S_{N-1}+\ldots\}+\ldots\,.
$$
By comparing the coefficients before $S_N$ we have
$$
\mathrm{e}^{(N\alpha+\beta)\tau}\{\mu_N+A(c_{N+1}\alpha+d_{N+1}\beta)\}=D(\mu_N)\mathrm{e}^{((N-1)\alpha+\beta)\tau}
$$
It follows that $\mu_N$ is a constant and then
$$
A(c_{N+1}\alpha+d_{N+1}\beta)=\mu_N(\mathrm{e}^{-\alpha \tau}-1).
$$
If
$c_{N+1}\alpha+d_{N+1}\beta=N\left\{\frac{N-1}{2}\alpha+\beta\right\}\ne
0$, then
$$
A(\tau)=C_1(\mathrm{e}^{-\alpha\tau}-1)$$
for some constant $C_1$.\\
If
$c_{N+1}\alpha+d_{N+1}\beta=N\left\{\frac{N-1}{2}\alpha+\beta\right\}=
0$ (in this case $\mu_N=0$) we compare coefficients before
$S_{N-1}$ and have
$$
\mathrm{e}^{(N\alpha+\beta)\tau}\{\mu_{N-1}+A(p_{N+1}A'+q_{N+1}A)\}=D(\mu_{N-1})\mathrm{e}^{((N-2)\alpha+\beta)\tau}.
$$
It follows that $\mu_{N-1}$ is a constant and
$$
p_{N+1}AA'+q_{N+1}A^2 =\mu_{N-1}(\mathrm{e}^{-2\alpha\tau}-1).
$$
Note that if
$c_{N+1}\alpha+d_{N+1}\beta=N\left\{\frac{N-1}{2}\alpha+\beta\right\}=
0$ then $p_{N+1}=-\frac{N(N-1)(N+1)}{12}\alpha\ne 0$ and $
q_{N+1}=-\frac{(N-1)N(N+1)}{24}\alpha^2\ne 0$ for $N\geq 2$.
Therefore, $\frac{2}q_{N+1}{p_{N+1}}=\alpha$.  Case $N=1$ should
be studied separately ($S_2=\mu_1S_1+\mu_0S_0$) and it was
already. Let us solve the equation
$$
p_{N+1}AA'+q_{N+1}A^2=\mu_{N-1}(\mathrm{e}^{-2\alpha\tau}-1).
$$
Denote by $y=A^2$. We have,
$$
y'+\alpha y=k_1\mathrm{e}^{-2\alpha \tau}-k_1
$$
for some constant $k_1$. It follows that
$$
A^2(\tau)=K_1(\mathrm{e}^{-2\alpha\tau}+K_2\mathrm{e}^{-\alpha
\tau}+1)
$$
for some constants $K_1$ and $K_2$.

Construct new sequence of vector fields
$$S_2^*=[S_1, S_0], \quad S_{n+1}^*=[S_1, S_n^*], \quad n\geq 2.$$
Note that $S_2^*=-S_2$. Since $L_x$ is of finite dimension then
there exists number $M$ such that $S_0$, $S_1$, $\ldots$, $S_M^*$
are linearly independent and
$$
S_{M+1}^*=\mu_M^*S_M^*+\mu_{M-1}^*S_{M-1}^*+\ldots+\mu_0^*S_0.
$$
There are the following possibilities.
$$
1) \left\{ \begin{array}{l}
A(\tau)=K_1(\mathrm{e}^{-\alpha \tau }-1),\\
B(\tau)=K_3(\mathrm{e}^{-\beta \tau }-1),
\end{array}
\right.
$$
$$
2) \left\{ \begin{array}{l}
A(\tau)=K_1(\mathrm{e}^{-\alpha \tau }-1),\\
B^2(\tau)=K_3^2(\mathrm{e}^{-2\beta \tau }+K_4\mathrm{e}^{-\beta
\tau}+1), \quad
S_{M+1}^*=\mu_M^*S_M^*+\mu_{M-1}^*S_{M-1}^*+\ldots+\mu_0^*S_0,
\quad \frac{M-1}{2}\beta+\alpha=0,
\end{array}
\right.
$$
$$
3) \left\{ \begin{array}{l}
B(\tau)=K_3(\mathrm{e}^{-\beta \tau }-1),\\
A^2(\tau)=K_1^2(\mathrm{e}^{-2\alpha \tau }+K_2\mathrm{e}^{-\alpha
\tau}+1), \quad S_{N+1}=\mu_NS_N+\mu_{N-1}S_{N-1}+\ldots+\mu_0S_0,
\quad \frac{N-1}{2}\alpha+\beta=0,
\end{array}
\right.
$$
$$
4) \left\{ \begin{array}{l} A^2(\tau)=K_1^2(\mathrm{e}^{-2\alpha
\tau }+K_2\mathrm{e}^{-\alpha \tau}+1), \quad
S_{N+1}=\mu_NS_N+\mu_{N-1}S_{N-1}+\ldots+\mu_0S_0,
\quad \frac{N-1}{2}\alpha+\beta=0,\\
B^2(\tau)=K_3^2(\mathrm{e}^{-2\beta \tau }+K_4\mathrm{e}^{-\beta
\tau}+1), \quad
S_{M+1}^*=\mu_M^*S_M^*+\mu_{M-1}^*S_{M-1}^*+\ldots+\mu_0^*S_0,
\quad \frac{M-1}{2}\beta+\alpha=0,
\end{array}
\right.
$$
where $K_1$, $K_2\ne -2$, $K_3$, $K_4\ne -2$ are some constants,
$M, N\geq 2$.

In case 1), vector fields  $S_0$ and $S_1$ generate an infinite
dimensional Lie algebra $L_x$ unless $\alpha+\beta=0.$

In case 2), we make a substitution
$1-\mathrm{e}^{\alpha\tau}=\mathrm{e}^{-\alpha w}$. Vector fields
$S_0$ and $S_1$ become
$$
S_0=K_1\frac{\partial }{\partial w} +\ldots,
$$
$$
S_1=\{K_3^2((1-\mathrm{e}^{-\alpha
w})^{-\frac{2\beta}{\alpha}}+K_4(1-\mathrm{e}^{-\alpha
w})^{-\frac{\beta}{\alpha}}+1)\}^{1/2}\frac{\partial }{\partial w}
+\ldots=g(w)\frac{\partial }{\partial w} +\ldots\,.
$$
Note that if
$$
S_{M+1}^*=\mu_M^*S_M^*+\mu_{M-1}^*S_{M-1}^*+\ldots+\mu_0^*S_0,
$$ then all coefficients $\mu_k^*$ are constants. By comparing coefficints before $\frac{\partial }{\partial w}$ in both sides of the last equation   we obtain that $g(w)$ is a solution of linear differential equation with constant
coefficients, that is
\begin{equation}\label{square}
g(w)=\{K_3^2((1-\mathrm{e}^{-\alpha
w})^{-\frac{2\beta}{\alpha}}+K_4(1-\mathrm{e}^{-\alpha
w})^{-\frac{\beta}{\alpha}}+1)\}^{1/2}
=\sum\limits_{k}R_k(w)\mathrm{e}^{\nu_k w}\, ,
\end{equation}
where $R_k(w)$ are some polynomials. One can show that equality
(\ref{square}) holds only if $B(\tau)=K_3(\mathrm{e}^{\alpha\tau
}+1)$. It can  be shown that in case 3)
$A(\tau)=K_1(\mathrm{e}^{\beta \tau}+1)$. In case 4) we make
substitution $\mathrm{e}^{\alpha\tau}
+\frac{K_1}{2}+\sqrt{\mathrm{e}^{2\alpha\tau +
K_1\mathrm{e}^{\alpha \tau}+1}} =\mathrm{e}^{\alpha w}$. Then
\begin{eqnarray*}
S_0&=&K_1\frac{\partial }{\partial w} +\ldots,\\
S_1&=&\Big\{K_3^2\Big( \frac{1}{2}\mathrm{e}^{\alpha
w}-\frac{K_1}{2}+\Big(\frac{K_1^2}{8}-\frac{1}{2}\Big)\mathrm{e}^{-\alpha
w}\Big)^{-\frac{2\beta}{\alpha}}\\
&+&K_4\Big(\frac{1}{2}\mathrm{e}^{\alpha
w}-\frac{K_1}{2}+\Big(\frac{K_1^2}{8}-\frac{1}{2}\Big)\mathrm{e}^{-\alpha
w}\Big)^{-\frac{\beta}{\alpha}}+1\Big)\Big\}^{1/2}\frac{\partial
}{\partial w} +\ldots\\
&=&g(w)\frac{\partial }{\partial w} +\ldots
\end{eqnarray*}
For function $g(w)$ to be of the form
$\sum\limits_{k}R_k(w)\mathrm{e}^{\nu_k w}$, where $R_k(w)$ are
polynomials, function $B(\tau)$ has to be of the form
$B(\tau)=K_3(\mathrm{e}^{\alpha}+1)$. Then, by case 3),
$A(\tau)=K_1(\mathrm{e}^{-\alpha \tau }+1)$.

It has been  proved that in cases 1), 2), 3), 4) one has
$$
1^*) \left\{ \begin{array}{l}
A(\tau)=K_1(\mathrm{e}^{-\alpha \tau }-1),\\
B(\tau)=K_3(\mathrm{e}^{\alpha \tau }-1),
\end{array}
\right.
$$

$$
2^*) \left\{ \begin{array}{l}
A(\tau)=K_1(\mathrm{e}^{-\alpha \tau }-1),\\
B(\tau)=K_3(\mathrm{e}^{\alpha \tau }+1),
\end{array}
\right.
$$
$$
3^*) \left\{ \begin{array}{l} A(\tau)=K_1(\mathrm{e}^{-\alpha \tau }+1),\\
B(\tau)=K_3(\mathrm{e}^{\alpha \tau }-1),
\end{array}
\right.
$$
$$
4^*) \left\{ \begin{array}{l} A(\tau)=K_1(\mathrm{e}^{-\alpha \tau
}+1),
\\
B(\tau)=K_3(\mathrm{e}^{\alpha \tau }+1).
\end{array}
\right.
$$
In case $1^*)$ function $d(t,t_1)$ in (\ref{main}) has a form
$d(t,t_1)=c_4(e^{\alpha t_1}-e^{\alpha t})+c_5(e^{-\alpha
t_1}-e^{-\alpha t})$, where $c_4$ and $c_5$ are some constants.
Equation (\ref{main}) with such function $d(t,t_1)$ admits a
nontrivial $x$-integral (see Introduction, Theorem 3 and \S 8).

In the next two sections we show that Cases $3^*$) and $4^*$) both
correspond to infinite dimensional Lie algebra $L_x$. Case $2^*$)
also produces an infinite dimensional Lie algebra $L_x$. It can be
proved in the same way as it is proved for case $3^*$).

\section{Characteristic Lie Algebra $L_x$ of the chain
$$t_{1x}=t_x+A_1(\mathrm{e}^{\alpha t_1}+\mathrm{e}^{\alpha t})-A_2(\mathrm{e}^{-\alpha
t}-\mathrm{e}^{-\alpha t_1})$$}

\bigskip

Since $A(\tau)=A_1(\mathrm{e}^{-\alpha\tau}+1)$ and
$B(\tau)=A_2(\mathrm{e}^{\alpha\tau}-1)$ then
$$
A(\tau)\mathrm{e}^{\alpha t}+\sum_{j=1}^k
A(\tau_j)\mathrm{e}^{\alpha t_j}=A_1\Big(\mathrm{e}^{\alpha
t}+\Big(2\sum_{j=1}^{k-1}\mathrm{e}^{\alpha
t_j}\Big)+\mathrm{e}^{\alpha t_k}\Big),
$$
and
$$
B(\tau)\mathrm{e}^{-\alpha t}+\sum_{j=1}^k
B(\tau_j)\mathrm{e}^{-\alpha t_j}=A_2(\mathrm{e}^{-\alpha
t}-\mathrm{e}^{-\alpha t_k}).
$$
We have,
$$
\frac{1}{A_1}S_0= (\mathrm{e}^{\alpha t}+\mathrm{e}^{\alpha
t_1})\frac{\partial }{\partial t_1} +\sum_{k=1}^\infty
 \Big(\mathrm{e}^{\alpha
t}+\Big(2\sum_{j=1}^{k-1}\mathrm{e}^{\alpha
t_j}\Big)+\mathrm{e}^{\alpha t_k}\Big)\frac{\partial }{\partial
t_k}+\sum_{k=1}^\infty
 \Big(\mathrm{e}^{\alpha
t}+\Big(2\sum_{j=1}^{k-1}\mathrm{e}^{\alpha
t_{-j}}\Big)+\mathrm{e}^{\alpha t_{-k}}\Big)\frac{\partial
}{\partial t_{-k}},
$$
and
$$
\frac{1}{A_2}S_1=\mathrm{e}^{-\alpha \tau}
\tilde{X}-\sum_{k=-\infty}^\infty \mathrm{e}^{-\alpha
t_k}\frac{\partial }{\partial t_k}=\mathrm{e}^{-\alpha
\tau}\tilde{X}-\tilde{S}_1,
$$
where $$
 \tilde{S}_1=\sum_{k=-\infty}^\infty
\mathrm{e}^{-\alpha t_k}\frac{\partial }{\partial t_k}.
$$
In variables $w_j=\frac{1}{\alpha}\mathrm{e}^{\alpha t_j}$ vector
fields $\tilde{S}_1$ and $\frac{1}{A_1}S_0$ can be rewritten as
$$
\tilde{S}_1=\sum\limits_{k=-\infty}^\infty \frac{\partial
}{\partial w_j},
$$
$$
\frac{1}{A_1}S_0=\alpha^2\sum\limits_{k=1}^\infty
\{w_k(w+2\sum\limits_{j=1}^{k-1}w_j)+w_k^2\}\frac{\partial}{\partial
w_k}+ \alpha^2\sum\limits_{k=1}^\infty
\{w_{-k}(w+2\sum\limits_{j=1}^{k-1}w_{-j})+w_{-k}^2\}\frac{\partial}{\partial
w_{-k}}.
$$
We have
$$
T_1=[\tilde{S}_1, [\tilde{S}_1, \frac{1}{\alpha^2
A_1}S_0]]=4\sum\limits_{k=-\infty}^\infty k\frac{\partial
}{\partial w_k}=4\tilde{T}_1, \quad
\tilde{T}_1=\sum\limits_{k=-\infty}^\infty k\frac{\partial
}{\partial w_k},
$$
$$
T_2=[\tilde{S}_1, [\tilde{T}_1, \frac{1}{\alpha^2
A_1}S_0]]=3\sum\limits_{k=1}^\infty \{k^2-k+1\}(\frac{\partial
}{\partial w_k}+\frac{\partial }{\partial
w_{-k}})=3\tilde{T}_2-3\tilde{T}_1+3\tilde{S}_1, \quad
\tilde{T}_2=\sum\limits_{k=-\infty}^\infty k^2\frac{\partial
}{\partial w_k}.
$$
Assume that $\tilde{T}_m=\sum\limits_{k=-\infty}^\infty
k^m\frac{\partial }{\partial w_k}$, $m=1,2 \ldots, n$, are vector
fields from $L_x$. Then
\begin{eqnarray*}
T_{m+1}&=&[\tilde{S}_1, [\tilde{T}_m,
\frac{1}{\alpha^2A_1}S_0]]=\sum\limits_{k=1}^\infty\{2(1+2^m+3^m+\ldots+k^m)+2k^{m+1}-k^m\}
\Big(\frac{\partial }{\partial w_k}+\frac{\partial }{\partial
w_{-k}}\Big)\\
&=&\sum\limits_{k=1}^\infty
\Big\{2\Big(\frac{k^{m+1}}{m+1}+d_{m,m+1}k^m+\ldots+d_{1,
m+1}k+d_{0, m+1}\Big)+2k^{m+1}-k^m\Big\}\Big(\frac{\partial
}{\partial w_k}+\frac{\partial }{\partial w_{-k}}\Big)
\end{eqnarray*}
and therefore, $\tilde{T}_{m+1}=\sum\limits_{k=-\infty}^\infty
k^{m+1}\frac{\partial }{\partial w_k}\in L_x$. It shows that
$\tilde{T}_n=\sum\limits_{k=-\infty}^\infty k^n\frac{\partial
}{\partial w_k}\in L_x$ for all $n=1,2,3,\ldots$, and $L_x$ is of
infinite dimension.

\section{Characteristic Lie Algebra $L_x$ of the chain
$$t_{1x}=t_x+A_1(\mathrm{e}^{\alpha t_1}+\mathrm{e}^{\alpha t})+A_2(\mathrm{e}^{-\alpha
t}+\mathrm{e}^{-\alpha t_1})$$}

\bigskip

It was observed in previous studies (see, for instance,
\cite{LeznovSmirnovShabat}) that S-integrable models have the
characteristic Lie algebra of finite growth. The chain studied in
this section can easily be reduced to the semi-discrete
sine-Gordon model $t_{1x}=t_x+\sin t+\sin t_1$ which belongs to
the S-integrable class. It is remarkable that its algebra $L_x$ is
of finite growth. Or, more exactly, the dimension of the linear
space of multiple commutators grows linearly with the
multiplicity. Below we prove that the linear space $V_n$ of all
commutators of multiplicity $\leq n$ has a basis $\{
P_1,P_2,P_3,...P_{2k};Q_2,Q_4,...Q_{2k}\}$ for $n=2k$ and a basis
$\{ P_1,P_2,P_3,...P_{2k+1};Q_2,Q_4,...Q_{2k}\}$ for $n=2k+1$,
where the operators $P_j$ and $Q_j$ are defined consecutively
$$
\begin{array}{ll}P_1=[S_0,S_1]+\alpha S_0+\alpha S_1,
\qquad & Q_1=P_1,\\
P_2=[S_1,P_1], \qquad & Q_2=[S_0,Q_1],\\
P_3=[S_0,P_2]+\alpha P_2,\qquad  & Q_3=[S_1,Q_2]-\alpha Q_2,\\
P_{2n}=[S_1,P_{2n-1}],\qquad  & Q_{2n}=[S_0,Q_{2n-1}],\\
P_{2n+1}=[S_0,P_{2n}]+\alpha P_{2n}, \qquad &
Q_{2n+1}=[S_1,Q_{2n}]-\alpha Q_{2n},
\end{array}
$$
for $n\geq 1$. Direct calculations show that
\begin{eqnarray}\label{conjugates}
DP_1D^{-1}&=&P_1-2\alpha (S_0+S_1),\nonumber\\
DP_2D^{-1}&=&\mathrm{e}^{-\alpha\tau}(P_2+2\alpha P_1-2\alpha^2
(S_0+S_1)),\nonumber\\
DP_3D^{-1}&=&P_3+2\alpha Q_2-2\alpha
P_2-4\alpha^2P_1+4\alpha^3(S_0+S_1),\nonumber\\
DP_4D^{-1}&=&\mathrm{e}^{-\alpha\tau}(P_4+2\alpha Q_3-4\alpha^2
P_2+4\alpha^2 Q_2-4\alpha^3 P_1+4\alpha^4 (S_0+S_1)),\nonumber\\
DQ_2D^{-1}&=&\mathrm{e}^{\alpha\tau}(Q_2-2\alpha P_1+2\alpha^2(S_0+S_1)),\nonumber\\
DQ_3D^{-1}&=&Q_3+2\alpha Q_2-2\alpha P_2-4\alpha^2
P_1+4\alpha^3(S_0+S_1),\nonumber\\
DQ_4D^{-1}&=& \mathrm{e}^{\alpha\tau}(Q_4-2\alpha
P_3+2\alpha^2(P_2-Q_2)+4\alpha^3 P_1-4\alpha^4 (S_0+S_1)),\nonumber\\
P_3=Q_3&,&[S_1,P_2]=-\alpha P_2, [S_0,Q_2]=\alpha Q_2, [S_1,P_4]=-\alpha P_4, [S_0,Q_4]=\alpha Q_4.\nonumber\\
&&
\end{eqnarray}
The coefficient before $\displaystyle \frac{\partial}{\partial
\tau}$ in all vector fields $DP_iD^{-1}$, $DQ_iD^{-1}$, $1\leq i
\leq 4$ is zero.
\begin{lemma}
For $n\geq 1$ we have,
\begin{enumerate}
\item[\rm{(1)}] $DP_{2n+1}D^{-1}+2\alpha
\mathrm{e}^{\alpha\tau}DP_{2n}D^{-1}=P_{2n+1}+2\alpha Q_{2n}$,\\
\item[\rm{(2)}] $\mathrm{e}^{\alpha\tau} DP_{2n+2}D^{-1}-\alpha
DP_{2n+1}D^{-1}=P_{2n+2}+\alpha Q_{2n+1}$,\\
\item[\rm{(3)}] $DQ_{2n+1}D^{-1}-2\alpha
\mathrm{e}^{-\alpha\tau}DQ_{2n}D^{-1}=Q_{2n+1}-2\alpha P_{2n}$,\\
\item[\rm{(4)}] $\mathrm{e}^{-\alpha\tau}DQ_{2n+2}D^{-1}+\alpha
DQ_{2n+1}D^{-1}=Q_{2n+2}-\alpha P_{2n+1}$,\\
\item[\rm{(5)}] $P_{2n+1}=Q_{2n+1}$,\\
\item[\rm{(6)}] $[S_1,P_{2n+2}]=-\alpha P_{2n+2}$,\\
\item[\rm{(7)}] $[S_0,Q_{2n+2}]=\alpha Q_{2n+2}$.
\end{enumerate}
Moreover, the coefficient before $\displaystyle
\frac{\partial}{\partial \tau}$ in all vector fields $DP_kD^{-1}$,
$DQ_kD^{-1}$ is zero.
\end{lemma}
\noindent {\bf Proof}. We prove the Lemma by
induction on $n$. It follows from (\ref{conjugates}) that the base
of induction holds for $n=1$. Assume $(1)-(7)$ are
true for all $n$, $1\leq n\leq k$. Let us prove that $(1)$ is true
for $n=k+1$.
\begin{eqnarray*}
&&DP_{2n+3}D^{-1}=D([S_0,P_{2n+2}]+\alpha
P_{2n+2})D^{-1}=[\mathrm{e}^{\alpha\tau}S_0,DP_{2n+2}D^{-1}]+\alpha
DP_{2n+2}D^{-1}\\
&&=[\mathrm{e}^{\alpha\tau}S_0,\alpha
\mathrm{e}^{-\alpha\tau}DP_{2n+1}D^{-1}+\mathrm{e}^{-\alpha\tau}P_{2n+2}+\alpha
\mathrm{e}^{-\alpha\tau}Q_{2n+1}]+\alpha DP_{2n+2}D^{-1}\\
&&=-\alpha^2(1+\mathrm{e}^{-\alpha\tau})DP_{2n+1}D^{-1}+\alpha
\mathrm{e}^{-\alpha\tau}[\mathrm{e}^{\alpha\tau}
S_0,DP_{2n+1}D^{-1}]-\alpha(1+\mathrm{e}^{-\alpha\tau})P_{2n+2}\\
&&\quad
-\alpha^2(1+\mathrm{e}^{-\alpha\tau})Q_{2n+1}+P_{2n+3}-\alpha
P_{2n+2}+ \alpha Q_{2n+2}+\alpha DP_{2n+2}D^{-1}\\
&&=-\alpha^2 (1+\mathrm{e}^{-\alpha\tau})DP_{2n+1}D^{-1}+\alpha
\mathrm{e}^{-\alpha\tau}D[S_0,Q_{2n+1}]D^{-1}-\alpha(2+\mathrm{e}^{-\alpha\tau})P_{2n+2}
\\
&&\quad
-\alpha^2(1+\mathrm{e}^{-\alpha\tau})Q_{2n+1}+P_{2n+3}+\alpha
Q_{2n+2}+\alpha DP_{2n+2}D^{-1}\\
&&=-\alpha^2(1+\mathrm{e}^{-\alpha\tau})DP_{2n+1}D^{-1}+\alpha
Q_{2n+2}-\alpha^2P_{2n+1}-\alpha^2DQ_{2n+1}D^{-1}
-\alpha(2+\mathrm{e}^{-\alpha\tau})P_{2n+2}\\
&&\quad -\alpha^2(1+\mathrm{e}^{-\alpha\tau})Q_{2n+1}-2\alpha^2
Q_{2n+1}-2\alpha P_{2n+2}+P_{2n+3}\\
&&=-2\alpha^2 DP_{2n+1}D^{-1}+2\alpha Q_{2n+2}-2\alpha^2
Q_{2n+1}-2\alpha P_{2n+2}+P_{2n+3}\\
&&=2\alpha P_{2n+2}+2\alpha^2 Q_{2n+1}-2\alpha
\mathrm{e}^{\alpha\tau}DP_{2n+2}D^{-1}+2\alpha Q_{2n+2}-2\alpha^2
Q_{2n+1}-2\alpha P_{2n+2}+P_{2n+3}\\
&&=-2\alpha \mathrm{e}^{\alpha\tau}DP_{2n+2}D^{-1}+2\alpha
Q_{2n+2}+P_{2n+3}.
\end{eqnarray*}
The proof of $(3)$ is the same as the proof of $(1)$. Let us show
that $(5)$ is true for $n=k+1$. We have,
\begin{eqnarray*}
DP_{2n+3}D^{-1}&=&-2\alpha
\mathrm{e}^{\alpha\tau}DP_{2n+2}D^{-1}+2\alpha
Q_{2n+2}+P_{2n+3}\\
&=&-2\alpha (\alpha DP_{2n+1}D^{-1}+P_{2n+2}+\alpha
Q_{2n+1})+2\alpha Q_{2n+2}+P_{2n+3},
\end{eqnarray*}
and
\begin{eqnarray*}
DQ_{2n+3}D^{-1}&=&2\alpha
\mathrm{e}^{-\alpha\tau}DQ_{2n+2}D^{-1}-2\alpha
P_{2n+2}+Q_{2n+3}\\
&=&2\alpha (-\alpha DQ_{2n+1}D^{-1}+Q_{2n+2}-\alpha
P_{2n+1})-2\alpha P_{2n+2}+Q_{2n+3}.
\end{eqnarray*}
By $(5)$, $P_{2n+1}=Q_{2n+1}$ and therefore
\begin{equation*}
D(P_{2n+3}-Q_{2n+3})D^{-1}=-2\alpha P_{2n+2}-2\alpha
Q_{2n+2}+2\alpha Q_{2n+2}+2\alpha P_{2n+2}=0.
\end{equation*}
Hence, $P_{2n+3}=Q_{2n+3}$.\\

\noindent Let us prove $(2)$ is true for $n=k+1$. We have,
\begin{eqnarray*}
&&\mathrm{e}^{\alpha\tau}DP_{2n+1}D^{-1}=\mathrm{e}^{\alpha\tau}D[S_1,P_{2n+3}]D^{-1}=\mathrm{e}^{\alpha\tau}[\mathrm{e}^{-\alpha\tau}S_1,DP_{2n+3}D^{-1}]\\
&&=\mathrm{e}^{\alpha\tau}[\mathrm{e}^{-\alpha\tau}S_1,-2\alpha
\mathrm{e}^{\alpha\tau}DP_{2n+2}D^{-1}+2\alpha Q_{2n+2}+P_{2n+3}]\\
&&=\mathrm{e}^{\alpha\tau}(-2\alpha^2(1+\mathrm{e}^{\alpha\tau})DP_{2n+2}D^{-1})
-2\alpha
\mathrm{e}^{2\alpha\tau}[\mathrm{e}^{-\alpha\tau}S_1,DP_{2n+2}D^{-1}]+P_{2n+4}+2\alpha
Q_{2n+3}+2\alpha^2 Q_{2n+2}\\
&&=-2\alpha^2(\mathrm{e}^{\alpha\tau}+\mathrm{e}^{2\alpha\tau})DP_{2n+2}D^{-1}+2\alpha^2\mathrm{e}^{2\alpha\tau}DP_{2n+2}D^{-1}+P_{2n+4}
+2\alpha Q_{2n+3}+2\alpha^2 Q_{2n+2}\\
&&=-2\alpha^2\mathrm{e}^{\alpha\tau}DP_{2n+2}D^{-1}+P_{2n+4}+2\alpha
Q_{2n+3}+2\alpha^2 Q_{2n+2}\\
&&=\alpha DP_{2n+3}D^{-1}-\alpha P_{2n+3}-2\alpha^2
Q_{2n+2}+P_{2n+4}+2\alpha Q_{2n+3}+2\alpha^2 Q_{2n+2}\\
&&=\alpha DP_{2n+3}D^{-1}+\alpha Q_{2n+3}+P_{2n+4}.
\end{eqnarray*}
The proof of $(4)$ is similar to the proof of $(2)$.\\

\noindent Let us prove that $(6)$ is true for $n=k+1$.
\begin{eqnarray*}
&&D[S_1,P_{2n+4}]D^{-1}=[\mathrm{e}^{-\alpha\tau}S_1,\alpha
\mathrm{e}^{-\alpha\tau}DP_{2n+3}D^{-1}+\mathrm{e}^{-\alpha\tau}P_{2n+4}+\alpha
\mathrm{e}^{-\alpha\tau}Q_{2n+3}]\\
&&=[\mathrm{e}^{-\alpha\tau}S_1,\alpha
\mathrm{e}^{-\alpha\tau}(-2\alpha
\mathrm{e}^{\alpha\tau}DP_{2n+2}D^{-1}+P_{2n+3}+2\alpha
Q_{2n+2})+\mathrm{e}^{-\alpha\tau}P_{2n+4}+\alpha \mathrm{e}^{-\alpha\tau}Q_{2n+3}]\\
&&=[\mathrm{e}^{-\alpha\tau}S_1,-2\alpha^2DP_{2n+2}D^{-1}+2\alpha
\mathrm{e}^{-\alpha\tau}P_{2n+3}+2\alpha^2\mathrm{e}^{-\alpha\tau}Q_{2n+2}+\mathrm{e}^{-\alpha\tau}P_{2n+4}]\\
&&=-2\alpha^2D[S_1,P_{2n+2}]D^{-1}-2\alpha^2\mathrm{e}^{-2\alpha\tau}(1+\mathrm{e}^{\alpha\tau})P_{2n+3}
-2\alpha^3\mathrm{e}^{-2\alpha\tau}(1+\mathrm{e}^{\alpha\tau})Q_{2n+2}\\
&&\quad+2\alpha
\mathrm{e}^{-2\alpha\tau}P_{2n+4}+2\alpha^2\mathrm{e}^{-2\alpha\tau}Q_{2n+3}+2\alpha^3\mathrm{e}^{-2\alpha\tau}Q_{2n+2}
-\alpha \mathrm{e}^{-2\alpha\tau}(1+\mathrm{e}^{\alpha\tau})P_{2n+4}+\mathrm{e}^{-2\alpha\tau}[S_1,P_{2n+4}]\\
&&=2\alpha^3DP_{2n+2}D^{-1}-2\alpha^2\mathrm{e}^{-\alpha\tau}P_{2n+3}+\alpha(\mathrm{e}^{-2\alpha\tau}-\mathrm{e}^{-\alpha\tau})P_{2n+4}
-2\alpha^3\mathrm{e}^{-\alpha\tau}Q_{2n+2}+\mathrm{e}^{-2\alpha\tau}[S_1,P_{2n+4}]\\
&&=\alpha^2\mathrm{e}^{-\alpha\tau}P_{2n+3}+2\alpha^3\mathrm{e}^{-\alpha\tau}Q_{2n+2}-\alpha^2\mathrm{e}^{-\alpha\tau}DP_{2n+3}D^{-1}
-2\alpha^2\mathrm{e}^{-\alpha\tau}P_{2n+3}\\
&&\quad+\alpha(\mathrm{e}^{-2\alpha\tau}-\mathrm{e}^{-\alpha\tau})P_{2n+4}
-2\alpha^3\mathrm{e}^{-\alpha\tau}Q_{2n+2}+\mathrm{e}^{-2\alpha\tau}[S_1,P_{2n+4}]\\
&&=-\alpha^2\mathrm{e}^{-\alpha\tau}P_{2n+3}+\alpha(\mathrm{e}^{-2\alpha\tau}-\mathrm{e}^{-\alpha\tau})P_{2n+4}
-\alpha DP_{2n+4}D^{-1}+\alpha
\mathrm{e}^{-\alpha\tau}P_{2n+4}\\
&&\quad+\alpha^2\mathrm{e}^{-\alpha\tau}Q_{2n+3}+\mathrm{e}^{-2\alpha\tau}[S_1,P_{2n+4}].
\end{eqnarray*}
Thus,
\begin{eqnarray*}
&&D[S_1,P_{2n+4}]D^{-1}=\mathrm{e}^{-2\alpha\tau}[S_1,P_{2n+4}]+\alpha
\mathrm{e}^{-2\alpha\tau}P_{2n+4}-\alpha DP_{2n+4}D^{-1}\\
&&D([S_1,P_{2n+4}]+\alpha
P_{2n+4})D^{-1}=\mathrm{e}^{-2\alpha\tau}([S_1,P_{2n+4}]+\alpha
P_{2n+4}).
\end{eqnarray*}
Hence, $[S_1,P_{2n+4}]=-\alpha P_{2n+4}$. $\Box$\\

\noindent Proof of $(7)$ is similar to the proof of $(6)$.

\begin{cor}\label{longcorollary} We have,
\begin{eqnarray*}
&&\mathrm{e}^{-\alpha\tau}DQ_{2n}D^{-1}+\mathrm{e}^{\alpha\tau}DP_{2n}D^{-1}=Q_{2n}+P_{2n}, \\
&&DP_{2n+1}D^{-1}=P_{2n+1}+\displaystyle
\sum_{k=1}^{n}(\mu_{2k}^{(2n+1)}P_{2k}+\nu_{2k}^{(2n+1)}Q_{2k})+
\sum_{k=0}^{n-1}\mu_{2k+1}^{(2n+1)}P_{2k+1}+\mu_0^{(2n+1)}S_0+\nu_0^{(2n+1)}S_1, \\
&&\displaystyle
DP_{2n}D^{-1}=\mathrm{e}^{-\alpha\tau}(P_{2n}+\sum_{k=1}^{n-1}(\mu_{2k}^{(2n)}P_{2k}+\nu_{2k}^{(2n)}Q_{2k})
+\sum_{k=0}^{n-1}\mu_{2k+1}^{(2n)}P_{2k+1}+\mu_0^{(2n)}S_0+\nu_0^{(2n)}S_1), \\
&&\displaystyle
DQ_{2n}D^{-1}=\mathrm{e}^{\alpha\tau}(Q_{2n}-\sum_{k=1}^{n-1}(\mu_{2k}^{(2n)}P_{2k}+\nu_{2k}^{(2n)}Q_{2k})
-\sum_{k=0}^{n-1}\mu_{2k+1}^{(2n)}P_{2k+1}-\mu_0^{(2n)}S_0-\nu_0^{(2n)}S_1).
\end{eqnarray*}
Moreover, $\mu_{2n}^{(2n+1)}=-2\alpha$,
$\nu_{2n}^{(2n+1)}=2\alpha$, $\mu_{2n-1}^{(2n)}=2\alpha$.
\end{cor}

\noindent Assume $L_x$ is of finite dimension. There are three
possibilities:
\begin{enumerate}
\item[1)]$S_0, S_1, P_1, P_2, Q_2, P_3, P_4, Q_4,..., P_{2n-1}$
are linearly independent and\\
$S_0, S_1, P_1, P_2, Q_2, P_3, P_4, Q_4,..., P_{2n-1}, P_{2n}$ are
linearly dependent,\\
\item[2)]$S_0, S_1, P_1, P_2, Q_2, P_3, P_4, Q_4,..., P_{2n-1},
P_{2n}$
are linearly independent and\\
$S_0, S_1, P_1, P_2, Q_2, P_3, P_4, Q_4,..., P_{2n-1}, P_{2n},
Q_{2n}$ are
linearly dependent,\\
\item[3)]$S_0, S_1, P_1, P_2, Q_2, P_3, P_4, Q_4,..., P_{2n},
Q_{2n}$
are linearly independent and\\
$S_0, S_1, P_1, P_2, Q_2, P_3, P_4, Q_4,..., P_{2n}, Q_{2n},
P_{2n+1}$ are linearly dependent.
\end{enumerate}
\noindent In case $1)$,
\begin{equation*}
P_{2n}=\gamma_{2n-1}P_{2n-1}+\gamma_{2n-2}P_{2n-2}+\eta_{2n-2}Q_{2n-2}+...
\end{equation*}
and
\begin{equation}\label{conjugateP_2n}
DP_{2n}D^{-1}=D(\gamma_{2n-1})DP_{2n-1}D^{-1}+D(\gamma_{2n-2})DP_{2n-2}D^{-1}+D(\eta_{2n-2})DQ_{2n-2}D^{-1}+...\,.
\end{equation}
We use Corollary \ref{longcorollary} to compare the coefficients
before $P_{2n-1}$ in (\ref{conjugateP_2n}) and have the
contradictory equality,
\begin{equation*}
\mathrm{e}^{-\alpha\tau}(\gamma_{2n-1}+2\alpha)=D(\gamma_{2n-1}).
\end{equation*}
It shows that case $1)$ is impossible to have.\\

\noindent In case $2)$,
\begin{equation*}
Q_{2n}=\gamma_{2n}P_{2n}+\gamma_{2n-1}P_{2n-1}+\eta_{2n-2}Q_{2n-2}+...
\end{equation*}
and
\begin{equation}\label{conjugateQ_2n}
DQ_{2n}D^{-1}=D(\gamma_{2n})DP_{2n}D^{-1}+D(\gamma_{2n-1})DP_{2n-1}D^{-1}+D(\eta_{2n-2})DQ_{2n-2}D^{-1}+...\,.
\end{equation}
We use Corollary \ref{longcorollary} to compare the coefficients
before $P_{2n-1}$ in (\ref{conjugateQ_2n}) and have the
contradictory equation,
\begin{equation*}
\mathrm{e}^{\alpha\tau}(\gamma_{2n-1}-2\alpha)=D(\gamma_{2n-1}).
\end{equation*}
It shows that case $2)$ is impossible to have.\\

\noindent In case $3)$,
\begin{equation*}
P_{2n+1}=\eta_{2n}Q_{2n}+\gamma_{2n}P_{2n}+...
\end{equation*}
and
\begin{equation}\label{conjugateP_2n+1}
DP_{2n+1}D^{-1}=D(\eta_{2n})DQ_{2n}D^{-1}+D(\gamma_{2n})DP_{2n}D^{-1}+...\,.
\end{equation}
We use Corollary \ref{longcorollary} to compare the coefficients
before $P_{2n}$ in (\ref{conjugateP_2n+1}) and have the
contradictory equation,
\begin{equation*}
(\gamma_{2n}-2\alpha)=D(\gamma_{2n})\mathrm{e}^{-\alpha\tau}.
\end{equation*}
It shows that case $3)$ also fails to be true. Therefore,
characteristic Lie algebra $L_x$ is of infinite dimension.

\section{Finding x-integrals}

Now we are ready to prove the main Theorem 3, formulated in
Introduction. Really, in the previous sections we proved that if
chain (\ref{main}) admits a nontrivial $x$-integral then it is one
of the forms $(1)-(4)$. The list $i)-iv)$ allows one to prove the
inverse statement: each of the equations from the list admits
indeed a nontrivial $x$-integral. $\Box$

Let us explain briefly how we found the list $i)-iv)$. Since for
each equation $(1)-(4)$ we have constructed the related
characteristic Lie algebra to find $x$-integral $F$ one has to
solve the corresponding system of the first order partial
differential equations. Below we illustrate the method with the
case $(2)$, for which the basis of the characteristic algebra
$L_x$ is given by the vector fields
$$\tilde{Y}=\partial_x+Y_{a(\tau)t+b(\tau)},\quad T_1=Y_{-a(\tau)},\quad
\tilde{X}= \frac{\partial }{\partial t}+ \frac{\partial }{\partial
t_{1}}+ \frac{\partial  }{\partial t_{-1} }+ \frac{\partial
}{\partial t_{2} }+  \frac{\partial }{\partial t_{-2} }+\ldots\,
,$$ where $a(\tau)=c_0\tau$ and $b(\tau)=c_2\tau^2+c_3\tau$. Note
that $x$-integral $F$ of $(2)$ should satisfy the equations
$\tilde{Y}F=0$, $T_1F=0$ and $\tilde{X}F=0$. Introduce new
variables $t$,$w$,$w_{\pm1},\ldots $ where $w_j=\ln(\tau_j)$ and
$\tau_j=t_j-t_{j+1}.$ Vector fields $\tilde{X}$, $T_1$ and
$\tilde{Y}$ in new variables are rewritten as
\begin{eqnarray*}
\tilde{X}&=&\frac{\partial}{\partial t},\quad T_1=\sum_{j=
-\infty}^{\infty}c_0\frac{\partial}{\partial w_j},\\
\tilde{Y}&=&\frac{\partial}{\partial
x}-t\sum_{j=-\infty}^{\infty}c_0\frac{\partial}{\partial
w_j}+c_0\sum_{j=-\infty}^{\infty}\{\tilde{\rho}_j+\tilde{b}(w_j)\}\frac{\partial}{\partial
w_j}\\
&=&\frac{\partial}{\partial
x}-tT_1+c_0\sum_{j=-\infty}^{\infty}\{\tilde{\rho}_j+\tilde{b}(w_j)\}\frac{\partial}{\partial
w_j},
\end{eqnarray*}
where
$$
\tilde{\rho_j}=\left\{\begin{array}{cl}\sum\limits_{k=0}^{j-1}\mathrm{e}^{w_k},
&{\rm{if}}\quad j\geq 1;\\
0, & {\rm{if}}\quad j=0;\\
-\sum\limits_{k=j}^{-1}\mathrm{e}^{w_k}, & {\rm{if}}\quad j\leq
-1,
\end{array}
\right. \quad
\tilde{b}(w_j)=-\frac{1}{c_0}(c_2\mathrm{e}^{w_j}+c_3) \, .
$$
Note that since we have $\tilde{X}F=0$, $F$ does not depend on
$t$. Now let us consider the vector field
$$\tilde{Y}+tT_1=A=\frac{\partial}{\partial
x}+c_0\sum_{j=-\infty}^{\infty}
\{\tilde{\rho}_j+\tilde{b}(w_j)\}\frac{\partial}{\partial w_j}.$$
We can write the vector field $A$ explicitly as
\begin{eqnarray*}
A&=&\frac{\partial}{\partial
x}+\sum_{j=-\infty}^{\infty}\Big\{\Big(c_0\sum_{k=0}^{j-1}\mathrm{e}^{w_k}
\Big)-c_2\mathrm{e}^{w_j}-c_3\Big\}\frac{\partial}{\partial
w_j}\\
&=&\frac{\partial}{\partial
x}-\frac{c_3}{c_0}T_1+\sum_{j=-\infty}^{\infty}\Big\{\Big(c_0\sum_{k=0}^{j-1}\mathrm{e}^{w_k}\Big)-c_2\mathrm{e}^{w_j}\Big\}\frac{\partial}{\partial
w_j}.
\end{eqnarray*}
The commutator $[T_1,A]$ gives
$$[T_1,A]=c_0A-c_0\frac{\partial}{\partial x}+c_3 T_1.$$
Thus we have three vector fields
\begin{eqnarray*}
A-\frac{\partial}{\partial
x}+\frac{c_3}{c_0}T_1:=\tilde{A}&=&\sum_{j=-\infty}^{\infty}\Big\{\Big(c_0\sum_{k=0}^{j-1}\mathrm{e}^{w_k}\Big)-c_2\mathrm{e}^{w_j}\Big\}\frac{\partial}{\partial
w_j},\\
\frac{T_1}{c_0}:=\tilde{T_1}&=&\sum_{j=-\infty}^{\infty}\frac{\partial}{\partial
w_j},\quad \tilde{X}_1=\frac{\partial}{\partial x},
\end{eqnarray*}
that solve $\tilde{A}F=0$, $\tilde{T}_1F=0$, $\tilde{X}_1F=0$.
Note that $[\tilde{T}_1,\tilde{A}]=\tilde{A}$. Since
$\tilde{X}_1F=0$, $F$ does not depend on $x$. Hence we end up with
two equations. By Jacobi theorem the system of equations has a
nontrivial solution $F(w,w_1,w_2)$ depending on three variables.
Therefore we need first three terms of $\tilde{A}$ and
$\tilde{T}_1$;
\begin{eqnarray*}
\tilde{A}&=&-c_2w\frac{\partial}{\partial
w}+(c_0\mathrm{e}^{w}-c_2\mathrm{e}^{w_1})\frac{\partial}{\partial
w_1}+(c_0\mathrm{e}^{w}+c_0\mathrm{e}^{w_1}-c_2\mathrm{e}^{w_2})\frac{\partial}{\partial
w_2} ,\\
\tilde{T}_1&=&\frac{\partial}{\partial w}+\frac{\partial}{\partial
w_1}+\frac{\partial}{\partial w_2}\,.
\end{eqnarray*}
Now we again introduce new variables $w=\epsilon, \quad
w-w_1=\epsilon_1, \quad w_1-w_2=\epsilon_2. $ Vector fields
$\tilde{A}$ and $\tilde{T}_1$ in new variables are rewritten as
\begin{eqnarray*}
\tilde{A}=\mathrm{e}^{\epsilon}\Big\{-c_2\frac{\partial}{\partial
\epsilon}+((-c_2-c_0)
+c_2\mathrm{e}^{-\epsilon_1})\frac{\partial}{\partial
\epsilon_1}+((-c_2-c_0)\mathrm{e}^{-\epsilon_1}
+c_2\mathrm{e}^{-\epsilon_1-\epsilon_2})\frac{\partial}{\partial
\epsilon_2} \Big\},  \quad \tilde{T}_1=\frac{\partial}{\partial
\epsilon}.
\end{eqnarray*}
To find the $x$-integral {\it ii)} in Theorem 3  one has to solve
the equation
$$\Big\{((-c_2-c_0)
+c_2\mathrm{e}^{-\epsilon_1})\frac{\partial}{\partial
\epsilon_1}+\mathrm{e}^{-\epsilon_1}((-c_2-c_0)
+c_2\mathrm{e}^{-\epsilon_2})\frac{\partial}{\partial \epsilon_2}
\Big\}F=0.$$

\section {Conclusion}

In this article the problem of classification of Darboux
integrable nonlinear semi-discrete chains of hyperbolic type was
studied. An approach based on the notion of characteristic Lie
algebra was properly modified and successfully used. We gave a
complete list of hyperbolic type chains $t_{1x}=t_x+d(t,t_1)$
admitting nontrivial $x$-integrals. We demonstrated that the
method of characteristic Lie algebras provides an effective tool
to classify integrable discrete chains as well. The method did not
get much attention in the literature, to our knowledge there are
only two studies (see \cite{ShabatYamilov} and
\cite{zhiberMurtazina}) where the characteristic Lie algebras are
applied for solving the classification problem for the partial
differential equations and systems. Surprisingly first of them was
published in 1981 and the second one only twenty five years later.

\section*{Acknowledgments}
This work is partially supported by the Scientific and
Technological Research Council of Turkey (T\"{U}B{\.{I}}TAK). One
of the authors (IH) thanks also Russian Foundation for Basic
Research (RFBR) (grants $\#$ 06-01-92051KE-a and  $\#$
08-01-00440-a).


\begin{thebibliography}{EMG}

\bibitem[1]{AdlerStartsev} V. E. Adler, S. Ya. Startsev,
{\it On discrete analogues of the Liouville equation,} Teoret.
Mat. Fizika, \textbf{121}, no. 2, 271-284 (1999), (English
translation: Theoret. and Math. Phys. , \textbf{121}, no. 2,
1484-1495, (1999)).

 \bibitem[2]{Zabrodin}
A.V. Zabrodin, {\it Hirota differential equations} (In Russian),
Teor. Mat. Fiz., \textbf{113}, no. 2,  179-230 (1997), (English
translation: Theoret. and Math. Phys. , \textbf{113}, no. 2,
1347-1392 (1997)).

\bibitem[3]{Yamilov}
R. l. Yamilov, {\it Symmetries as integrability criteria for
differential difference equations,} J. Phys. A: Math. Gen. 39,
541-623 (2006).

\bibitem[4]{NijhoffCapel}
F. W. Nijhoff, H. W. Capel, {\it The discrete Korteweg-de Vries
equation}, Acta Applicandae Mathematicae, \textbf{39}, 133-158
(1995).

\bibitem[5]{GKP}
B. Grammaticos, G. Karra, V. Papageorgiou, A. Ramani, {\it
Integrability of discrete-time systems, Chaotic dynamics,}
(Patras,1991), NATO Adv. Sci. Inst. Ser. B Phys. , \textbf{298},
75-90, Plenum, New York, (1992).

\bibitem[6]{Darboux}
G. Darboux, {\it Le\c{c}ons sur la th$\acute{e}$orie
g$\acute{e}$n$\acute{e}$rale des surfaces et les applications
geometriques du calcul infinitesimal,} T.2. Paris: Gautier-Villars
(1915).

\bibitem[7]{AndersonKamran}
I. M. Anderson, N. Kamran, {\it The variational bicomplex for
hyperbolic second-order scalar partial differential equations in
the plane,} Duke Math. J. , \textbf{87}, no. 2, 265-319 (1997).


\bibitem[8]{Zhiber}  A. V. Zhiber,  V. V. Sokolov, {\it Exactly integrable
hyperbolic equations of Liouville type}, (In Russian) Uspekhi Mat.
Nauk 56, no. 1 (337), 63-106 (2001), (English translation: Russian
Math. Surveys, \textbf{56}, no. 1, 61-101 (2001)).

\bibitem[9]{ShabatYamilov}
A. B. Shabat, R. I. Yamilov, {\it Exponential systems of type I
and the Cartan matrices}, (In Russian), Preprint, Bashkirian
Branch of Academy of Science of the USSR, Ufa, (1981).

\bibitem[10]{LeznovSmirnovShabat}
A. N. Leznov, V. G. Smirnov, A. B. Shabat, {\it Group of inner
symmetries and integrability conditions for two-dimensional
dynamical systems,} Teoret. Mat. Fizika, \textbf{51}, no. 1, 10-21
(1982).

\bibitem[11]{ZhiberMukminov} A. V. Zhiber, F. Kh. Mukminov, {\it Quadratic systems,
symmetries, characteristic and complete algebras}, Problems of
Mathematical Physics and Asymptotics of their Solutions, ed.
L.A.Kalyakin, Ufa, Institute of Mathematics, RAN, 13-33 (1991).


\bibitem[12]{BormisovMukminov} A. A. Bormisov, F. Kh. Mukminov, {\it Symmetries of
hyperbolic systems of Riccati equation type}, (In Russian),
Teoret. Mat. Fiz. \textbf{127}, no. 1, 47-62 (2001), (English
translation: Theoret. and Math. Phys. , \textbf{127}, no. 1,
446-459 (2001)).

\bibitem[13]{SokolovStartsev} V.V. Sokolov and S.Ya. Startsev,
{\it Symmetries of nonlinear hyperbolic systems of the Toda chain
type}, Theoretical and Mathematical Physics, \textbf{155}, no. 2,
802-811 (2008).

\bibitem[14]{zhiberMurtazina}A. V. Zhiber, R. D. Murtazina, {\it On the
characteristic Lie algebras for the equations $u_{xy}=f(u,u_x)$},
(In Russian), Fundam. Prikl. Mat. , \textbf{12}, no. 7, 65-78
(2006).


\bibitem[15]{IbragimovShabat} N. Kh. Ibragimov, A. B. Shabat,
{\it Evolution equations with nontrivial Lie-B\"{a}cklund group,}
Funktsional. Anal. i Prilozhen, \textbf{14}, no. 1, 25-36 (1980).

\bibitem[16]{MSY} A. V. Mikhailov, A. B. Shabat, R. I. Yamilov,
{\it A symmetry approach to the classification of nonlinear
equations. Complete list of integrable systems}, (In Russian),
Uspekhi Mat. Nauk,  \textbf{42}, no. 4, 3-53 (1987).

\bibitem[17]{LeviYamilov}
R. I. Yamilov, D. Levi, {\it Integrability conditions for $\$n\$$
and $\$t\$$ dependent dynamical lattice equations}, J. Nonlinear
Math. Phys. , \textbf{11}, no. 1, 75-101 (2004).

\bibitem[18]{Gurses3}
M. G\"{u}rses, A. Karasu, {\it Integrable KdV Systems: Recursion
Operators of Degree Four,} Physics Letters A, \textbf{251},
247-249 (1999) // $\tt{arxiv: solv-int/9811013}$.

\bibitem[19]{Gurses4}
M. G\"{u}rses, A. Karasu, R. Turhan, {\it Nonautonomous Svinolupov
Jordan KdV Systems,} Journal of
 Physics A: Mathematical and General, \textbf{34}, 5705-5711 (2001) // $\tt{arxiv: nlin.SI/0101031}$.

\bibitem[20]{Svinolupov} S. I. Svinolupov, {\it On the analogues
of the Burgers Equation}, Phys. Lett. A, \textbf{135}, no. 1,
32-36 (1989).


\bibitem[21]{ZhiberShabat79} A. V. Zhiber, A. B. Shabat, {\it The
Klein-Gordon equation with nontrivial group}, (In Russian), Dokl.
Akad. Nauk USSR, \textbf{247}, no. 5, 1103-1107 (1979), (English
translation: Soviet Phys. Dokl. , \textbf{24}, 607-609 (1979)).

\bibitem[22]{ZhiberShabat84} A. V. Zhiber, A. B. Shabat, {\it Systems of
equations $u_x=p(u,v)$, $v_y=q(u,v)$ that possess symmetries}, (In
Russian), Dokl. Akad. Nauk USSR, \textbf{277}, no. 1, 29-33
(1984), (English translation: Soviet Math. Dokl. , \textbf{30},
23-26 (1984)).

\bibitem[23]{Habibullin}
I. T. Habibullin, {\it Characteristic algebras of fully discrete
hyperbolic type equations,} Symmetry, Integrability and Geometry:
Methods and Applications, no. 1, paper 023, 9 pages, (2005) //
$\tt{arxiv: nlin.SI/0506027}$.


\bibitem[24]{HabibullinPekcan}
I. Habibullin, A. Pekcan, {\it Characteristic Lie Algebra and
Classification of Semi-Discrete Models}, Theoret. and Math. Phys.,
\textbf{151}, no. 3, 781-790 (2007) //$\tt{ arXiv:nlin/0610074}$.

\bibitem[25]{TJM} I. Habibullin, N. Zheltukhina, A. Pekcan,
{\it On Some  Algebraic Properties of Semi-Discrete Hyperbolic
Type Equations,} Turkish Journal of Mathematics, 32, 1-17(2008)
//$\tt{ arXiv:nlin/0703065}$.




\end{thebibliography}
\end{document}